\documentclass[a4paper,journal]{IEEEtran}
\usepackage[T1]{fontenc}
\usepackage[utf8]{inputenc}
\usepackage{cite}
\usepackage{amsmath,amsfonts}
\usepackage{algorithm}
\usepackage{array}
\usepackage{subfig}
\usepackage{textcomp}
\usepackage{stfloats}
\usepackage{url}
\usepackage{verbatim}
\usepackage{graphicx}
\usepackage{booktabs}
\usepackage{amsthm}     
\usepackage{algorithm}	
\usepackage{algpseudocode}

\newtheorem{proposition}{Proposition}

\theoremstyle{definition}

\theoremstyle{remark}

\newtheorem{remark}{Remark}

\usepackage{amsmath}    
\usepackage{amsfonts}   
\usepackage{bm}         
\usepackage{multirow}	
\usepackage{times} 
\usepackage{enumitem}
\usepackage{graphicx}
\usepackage{xcolor}
\usepackage{setspace}
\usepackage{hyperref}

\DeclareMathOperator*{\argmax}{arg\,max}
\DeclareMathOperator*{\argmin}{arg\,min}

\allowdisplaybreaks

\usepackage{pgf}
\usepackage{tikz}
\usepackage{pgfplots}
\DeclareUnicodeCharacter{2212}{−}
\usepgfplotslibrary{groupplots, dateplot}
\usetikzlibrary{patterns,shapes.arrows}
\pgfplotsset{compat=newest}

\begin{document}

\title{Reinforcement Learning with Distributed MPC for Fuel-Efficient Platoon Control with Discrete Gear Transitions}
\newcommand{\mytitle}{Reinforcement Learning with Distributed MPC for Fuel Efficient Platoon Control with Discrete Gear Transitions}
\hypersetup{hidelinks,pdfauthor=Samuel Mallick, pdfcreator=Samuel Mallick, pdftitle=\mytitle}

\author{Samuel Mallick\textsuperscript{$\star$}, Gianpietro Battocletti\textsuperscript{$\star$}, Dimitris Boskos, \IEEEmembership{Member, IEEE}, \\Azita Dabiri, and Bart De Schutter, \IEEEmembership{Fellow, IEEE}
\thanks{\textsuperscript{$\star$} These authors contributed equally to this work.}
\thanks{This paper is part of a project that has received funding from the European Research Council (ERC) under the European Union's Horizon
	2020 research and innovation programme (Grant agreement No. 101018826
	- CLariNet).}
\thanks{All authors are affiliated with Delft Center for Systems and Control, Delft
	University of Technology, Delft, The Netherlands (e-mail: \{s.h.mallick, g.battocletti, d.boskos, a.dabiri, b.deschutter\}@tudelft.nl).}}



\maketitle

\begin{abstract}
\textcolor{black}{
Cooperative control of groups of autonomous vehicles (AVs), i.e., platoons, is a promising direction to improving the efficiency of autonomous transportation systems.
In this context, distributed co-optimization of both vehicle speed and gear position can offer benefits for fuel-efficient driving.
To this end, model predictive control (MPC) is a popular approach, optimizing the speed and gear-shift schedule while explicitly considering the vehicles' dynamics over a prediction window.
However, optimization over both the vehicles' continuous dynamics and discrete gear positions is computationally intensive, and may require overly long sample times or high-end hardware for real-time implementation.
This work proposes a reinforcement learning (RL)-based distributed MPC approach to address this issue.
For each vehicle in the platoon, a policy is trained to select and fix the gear positions across the prediction window of a local MPC controller, leaving a significantly simpler continuous optimization problem to be solved as part of a distributed MPC scheme.
In order to reduce the computational cost of training and facilitate the scalability of the proposed approach to large platoons, the policies are parameterized such that the emergent multi-agent RL problem can be decoupled into single-agent learning tasks.
In addition, a recurrent neural-network (RNN) architecture is proposed for the gear selection policy, such that the learning is scalable even as the number of possible gear-shift schedules grows exponentially with the MPC prediction horizon.
In highway-driving simulations, the proposed approach is shown to have a significantly lower computation burden and a comparable performance in terms of fuel-efficient platoon control, with respect to pure MPC-based co-optimization.
}
\end{abstract}


\section{Introduction}

\IEEEPARstart{A}{utonomous} \textcolor{black}{vehicles (AVs) have for many years been a tantalizing prospect for a practical demonstration of modern control and artificial intelligence techniques.
In particular, AVs have the potential to improve safety and to reduce traffic congestion, improving the efficiency of the transportation system in which they take part, while further improving local fuel economy through intelligent speed and powertrain management \cite{faisal2019understanding}.
When multiple AVs follow each other in a \emph{platoon}, e.g., in highway driving scenarios, these benefits can even be enhanced through coordinated driving patterns \cite{rebelo2024vehicle}.}

For control of platoons, key aspects include the communication topology, governing the exchange of information between AVs, the formation geometry, describing the desired coordinated driving pattern, and the distributed controllers, coordinating the individual behavior of AVs to achieve the global goal \cite{li2015overview}.
A variety of distributed controllers have been proposed in the literature.
These include optimization-based controllers \cite{lan2023datadriven, liu2019distributed}, where inter-vehicle communication is leveraged such that distributed optimization algorithms can be used to resolve a global optimal control problem online.
Additionally, learning-based controllers have been proposed \cite{chen2024communication, xu2024multi, li2021reinforcement}, in which control of the platoon is cast as a multi-agent learning problem, with each AV learning a control policy from driving data generated while interacting with the other AVs.
Beyond optimization and learning, alternative platoon controllers have been proposed, e.g., PID \cite{yang2020longitudinal}, sliding-mode control \cite{guo2023distributed}, and constraint-driven control \cite{beaver2022constraint}.
For a broader treatment of platoon control paradigms, see the survey \cite{lesch2022overview}.

Among the existing control strategies, model predictive control (MPC) is a powerful and prevalent approach \cite{mallick2024comparison, zheng2017distributed}, able to optimize each AV's trajectory while considering physical and safety-based constraints.
MPC controllers optimize a platoon's control inputs over a prediction window, explicitly considering the vehicles' dynamics, with the optimal inputs for the first predicted time step being applied to the system \cite{rawlings2017model}.
The optimization is then performed again at the next time step in a receding horizon fashion.
\textcolor{black}{
When the MPC optimization problem is a continuous nonlinear program (NLP), state-of-the-art numerical solvers are in general often able to resolve the optimization problem fast enough for an online implementation in reasonably-sized platoons \cite{wachter2006implementation}.
}
However, when considering AVs with step-gear transmissions, intelligent control of the vehicle's speed and gear-shift schedule is required in order to achieve the benefits of high-performing and fuel-efficient autonomous driving \cite{li2015eco}.
This introduces discrete inputs in the MPC optimization problem, which thus becomes a mixed-integer nonlinear program (MINLP) \cite{bemporad1999control}, for which the computational burden is, in general, intractable for real-time control.

To address this issue, in \cite{shao2021vehicle} and \cite{sun2022energy} it is proposed to simplify the MINLP to a mixed-integer quadratic program (MIQP) via approximating the  nonlinear motion and fuel consumption equations.
However, such approximations of the optimization problem can introduce significant suboptimality.
In \cite{ganesan2023numerical} the MINLP problem is solved approximately using heuristic numerical strategies.
However, there are no guarantees on the optimality gap introduced.
Similarly, in \cite{xu2015fuel} the problem is considered under the specific pulse-and-glide cruising strategy, where the MINLP is approximated by simple rules to make it computationally implementable.
In \cite{li2020ecological} the computational burden of the MINLP is alleviated by replacing the MPC controller completely with a fast learning-based controller that controls both speed and gear. However, this approach is not able to guarantee constraint satisfaction for, e.g., engine limits.
Furthermore, these works consider control of an individual AV following a prespecified driving pattern, e.g., set by a human-driven leading vehicle.
In contrast, when considering control of a cooperative platoon, the computational complexity and potential suboptimality are significantly larger. 

An alternative approach is to decouple the gear selection from control of the AV's continuous dynamics.
This approach is leveraged in \cite{he2018optimal}, where a PI controller first regulates the AV's speed, and a hybrid control problem is then solved to select the gears. However, again only a single AV is considered.
The decoupling approach has been explored also in the case of platoon control, where distributed optimization algorithms can optimize over the continuous dynamics in a cooperative way, with the gear selection of each AV then done locally once the speed is fixed.
In \cite{turri2016gear} dynamic programming is used to select a gear locally for each AV given the speed, while in \cite{yin2022hierarchical} the gears and engine torque are selected locally by a Q-learning controller, given a fixed speed that is optimized at the platoon level.
In general, however, decoupling the gear optimization from that of the continuous AV dynamics is suboptimal with respect to the co-optimization of both. 

Of particular relevance is the authors' previous work \cite{mallick2025learning}, in which a learning-based controller is trained to select a gear-shift \emph{schedule}, i.e., a sequence of gears across the prediction window of the MPC controller, approximating the discrete component of the solution to the MINLP, with the remaining optimization problem to be solved online then becoming an NLP.
However, this work is limited to the control of only a single vehicle.
Furthermore, the approach relies on supervised learning (SL), for which the generation of data can be computationally challenging, especially in the case of multiple vehicles in platoons.

In light of the above issues, this work presents a reinforcement learning (RL)-based MPC scheme for the co-optimization of speed and gear-shift schedule for a \emph{platoon} of AVs.
Taking inspiration from \cite{mallick2025learning, da2024integrating, mallick2024learning}, the discrete and continuous components of the MPC problem are isolated.
Learned policies select and fix the gear positions across the prediction window of local MPC controllers for each AV.
Optimal control and constraint satisfaction are then handled by a distributed MPC approach where nonlinear programs (NLPs) must be solved, rather than MINLPs. 
Rather than decoupling the gear choice from the speed control, the policies learn to select optimal gear-shift schedules across the prediction window of the MPC controllers.
Thus, the MPC controllers are able to consider the gear and powertrain dynamics without optimizing explicitly over discrete inputs.
In this way, while the computation of the gear-shift schedule and vehicle speed are decoupled, the notion of co-optimization is retained.
Furthermore, an RL approach is proposed to train the gear-shift schedule policies, with the policies' architecture allowing for training in a single vehicle scenario while still capturing inter-vehicle interaction, with the policies generalizing to the platoon case, thus avoiding the complexity of a multi-agent RL (MARL) approach.
Finally, feasibility of the MPC controllers is guaranteed by a heuristic gear-shift schedule that acts in parallel to the learning-based one. 

The contributions of this work are as follows:
\begin{itemize}
    \item With the computational burden of optimizing gear-shift schedules shifted offline, only NLPs must be solved online, yielding a lower online computational burden for the proposed approach with respect to approaches that solve mixed-integer optimization problems.
    \item With gear-shift schedules being selected by an RL policy, trained to generate optimal gear-shift schedules across the whole MPC prediction window, the closed-loop performance of the proposed approach outperforms approaches that decouple optimization of the vehicle speed and gear-shift schedule.
    \item A novel policy parameterization is proposed to address the MARL problem, allowing training in a single-vehicle scenario and generalization to the platoon case.
    Furthermore, in the case of homogeneous vehicles, a single policy can be trained and deployed in each of the vehicles.
    \item A recurrent architecture is proposed for the policy to address the exponential growth of the action space with the prediction horizon.
    As a further benefit of this structure, once trained, the policy generalizes over different prediction horizons.
\end{itemize}

The remainder of this paper is organized as follows.
In Section \ref{sec:background} the background material is introduced, as well as the problem setting.
In Section \ref{sec:mpc_controller} we introduce the distributed platoon controller.  Section \ref{sec:parameterized_mpc} introduces the computationally tractable learning-based MPC variant, with the learning component detailed in Section \ref{sec:policy}.
Section \ref{sec:results} validates the approach through experiments, and, finally, Section \ref{sec:conclusions} concludes the paper.

\section{Problem Setting and Background}
\label{sec:background}

\subsection{Preliminaries and notation}
We use superscripts to index the scalar elements of vectors, e.g., $x^{[1]} = a$ for $x = [a, b]^\top$. 
\textcolor{black}{
We define the clipping operator $\mathrm{clip}(x, a, b)$ that truncates each element $x^{[i]}$ of the vector $x=\big[x^{[1]}, \ldots ,x^{[n]}\big]^\top$ to lie within the range $[a, b]$, i.e., $\mathrm{clip}(x, a, b) = \big[y^{[1]}, \ldots, y^{[n]}\big]^\top$ with $y^{[i]}=x^{[i]}$ if $x^{[i]}\in[a,b]$, $y^{[i]}=a$ if $x^{[i]}\in(-\infty,a)$, and $y^{[i]}=b$ if $x^{[i]}\in(b,\infty)$.
}
As we consider a multi-AV setting, subscripts denote variables associated with a specific AV, e.g., $x_i$ is the variable $x$ for the $i$th AV.
Finally, when the specific AV is not relevant we occasionally drop the subscript, e.g., $x$ is for a general AV. 

\subsection{Reinforcement Learning and Deep Q-Network Algorithm}
\label{sec:rl}
In this section we introduce relevant background theory on RL that will be leveraged in the proposed approach.

Consider a discrete-time dynamical system modeled as a Markov Decision Process (MDP) \cite{sutton2018reinforcement} over a continuous state space $\mathcal{S}$, a discrete action space $\mathcal{A}$, a deterministic state transition function $F: \mathcal{S} \times \mathcal{A} \rightarrow \mathcal{S}$, and a stage cost $L : \mathcal{S} \times \mathcal{A} \to \mathbb{R}$ defining a cost for a given action in a given state.
Moreover, let $\pi_\theta: \mathcal{S} \rightarrow \mathcal{A}$ be a deterministic policy that for every state $s\in\mathcal{S}$ selects an action $a\in\mathcal{A}$ to apply.
Assuming an initial probability distribution over the states of the MDP, and selecting actions based on the deterministic policy $\pi_\theta$, yields a probability distribution over the MDPs state trajectories, denoted $\sigma_{\pi_\theta}$.
The performance of the policy is defined as \cite{sutton2018reinforcement}
\begin{eqnarray}
	P(\pi_\theta) = \mathbb{E}_{\sigma_{\pi_\theta}} \Bigg[\sum_{k=0}^\infty \gamma^k L\big(s_k, \pi_\theta(s_k)\big)\Bigg],
\end{eqnarray}
with $s_k \in \mathcal{S}$ the state at time step $k$, and $\gamma \in (0, 1]$ the discount factor.
The RL task is then to find the optimal policy $\pi_\theta^\ast$ as 
\begin{eqnarray}
	\pi_\theta^\ast = \argmin_\theta P(\pi_\theta).
\end{eqnarray}
As in this work we consider an RL problem with a discrete action space, we focus here on the Deep Q-Network (DQN) algorithm \cite{mnih2015human}, which is well suited for this type of problem.
As a value-based method, DQN considers the action-value function \cite{sutton2018reinforcement}, defined as
\begin{equation}
	Q_{\pi_\theta}(s_k, a_k) = L(s_k, a_k) + \mathbb{E}_{\sigma_{\pi_\theta}} \Bigg[\sum_{\tau=k+1}^\infty \gamma^{\tau-k} L\big(s_k, \pi_\theta(s_k)\big) \Bigg].
\end{equation}
In particular, DQN aims to approximate, with an neural-network (NN)-based approximator $Q_\theta$, the optimal action-value function 
\begin{eqnarray}
	Q_\theta \approx Q^\ast(s, a) = \min_{\pi_\theta} Q_{\pi_\theta} (s,a).
\end{eqnarray}
The policy is then inferred in a greedy manner 
\begin{eqnarray}
	\pi_\theta(s) = \argmin_a Q_\theta(s, a).
\end{eqnarray}
The DQN algorithm is summarized as follows \cite{mnih2015human}.
Two sets of weights $\theta$ and $\tilde{\theta}$ are maintained for a policy network $Q_\theta$ and a target network $Q_{\tilde{\theta}}$, respectively.
At each time step $k$, an action $a_k$ is chosen either via the policy or randomly, with the choice made based on an exploration probability $\epsilon(k)$.
The transition $\big(s_k, a_k, L(s_k, a_k), s_{k+1}\big)$ is stored in a finite buffer $\mathcal{D}$.
At each parameter update $N_\text{batch}$ transitions are sampled from the buffer, and the weights $\theta$ are updated via a gradient descent step on the loss function
\begin{equation}
	\label{eq:loss}
	\begin{aligned}
		L_{\text{loss}}(\theta) & = \sum_{l=1}^{N_\text{batch}} \ell \bigg(L(s^{(l)}_k, a^{(l)}_k) \\
		&+ \gamma \max_{a} Q_{\tilde{\theta}}\big(s^{(l)}_{k+1}, a\big) - Q_\theta\big(s^{(l)}_k, a^{(l)}_k\big)\bigg),
	\end{aligned}
\end{equation}
where $\ell$ is some appropriate penalty function, e.g., a smooth L1 loss.
Finally, the target network weights $\tilde{\theta}$ are updated with a blending, controlled by blending parameter $\nu \in (0, 1]$,
\begin{equation}
	\label{eq:target_blending}
	\tilde{\theta} \gets \nu \theta + (1-\nu)\tilde{\theta}.
\end{equation}

\subsection{AV model}
Consider the vehicle and powertrain models for an AV \cite{sun2014design}
\begin{equation}
	\label{eq:dynamics}
	\begin{aligned}
		T(t)n(t) &= G(t) + C v^2(t) + m a(t) + F(t), \\
		\omega(t) &= \frac{30}{\pi} \cdot n(t) v(t) ,
	\end{aligned}
\end{equation} 
with $t$ continuous time, $a$ the acceleration, $v$ the velocity, and $m$ the mass of the vehicle.
Furthermore, $C$ is the wind drag coefficient, $F$ is the brake force, $T$ is the engine torque, and $\omega$ is the engine speed.
The lumped gear ratio $n(t) = z\big(j(t)\big)z_\text{f}/r$ is determined by the final drive ratio $z_\text{f}$, the wheel radius $r$, and the transmission gear ratio $z$, which is a discrete variable selected by the gear position $j \in \{1,\dots,j_\mathrm{max}\}$, with $z(j+1) < z(j)$.
The friction function 
\begin{equation}
	G(t) = \mu m g \cos\big(\alpha(t)\big) + m g \sin\big(\alpha(t)\big), 
\end{equation}
with $\mu$ the rolling friction constant and $g$ the gravitational acceleration, defines the road friction for road angle $\alpha$, which, for simplicity of presentation, is assumed to be constant in the sequel, i.e., $\alpha(t) = \alpha$ and $G(t) = G$.
To capture the dynamics of the engine torque $T$, the rate of change of the torque is constrained as
\begin{equation}
	\label{eq:torque_rate}
	|\dot{T}(t)| \leq \Delta T_\text{max}.
\end{equation}
Furthermore, we consider the fuel consumption model \cite{shao2021vehicle}
\begin{equation}
	\label{eq:fuel}
	\dot{m}_{\text{f}}(t) = c^{[1]} + c^{[2]} \omega(t) + c^{[3]} \omega(t) T(t),
\end{equation}
with $c = \big[c^{[1]}, c^{[2]}, c^{[3]}\big]^\top$ a vector containing the model constants.

The variables $F,T,$ and $\omega$ are physically bounded above and below, e.g., $T_\text{min} \leq T \leq T_\text{max}$.
Note that the bounds on $\omega$ implicitly impose bounds on $v$ as:
\begin{equation}
	v_\text{min} = \frac{\pi \cdot \omega_\text{min} r}{30 \cdot z(1) z_\text{f}}, \quad \text{and}\quad v_\text{max} = \frac{\pi \cdot \omega_\text{max} r}{30 \cdot z(j_\mathrm{max}) z_\text{f}}.
\end{equation}
For convenience, in the following, we define a function $\omega$ that maps a speed and gear to the corresponding engine speed:
\begin{equation}
	\omega(v, j) = \frac{30 \cdot v \cdot z(j) z_\text{f}}{r \pi}.
\end{equation}

\subsection{Problem setting}
\label{sec:problem_setting}

Consider a platoon of $M$ AVs, potentially inhomogeneous via different values of model parameters, indexed with $i \in \mathcal{M} = \{1,\dots,M\}$, where, without loss of generality, the positions of the AVs are ordered as the indices of $\mathcal{M}$, i.e., for the first vehicle of the platoon (called the \emph{leader}) we have $i = 1$, while for the last AV we have $i = M$.
We consider the task of controlling the platoon of AVs in a fuel-efficient manner.
The leader of the platoon is tasked with tracking a reference trajectory, while all other AVs track the trajectory of their predecessor.
Denote the position of AV $i$, the reference position, and the reference velocity at time $t$ as $p_i(t)$, $p_\text{ref}(t)$, and $v_\text{ref}(t)$, respectively.
We consider then the following discrete-time performance metric for the task over a time interval of $K \Delta t$ time units:
\begin{equation}
	\label{eq:metric}
	\begin{aligned}
		&J(K) = \sum_{k = 0}^{K-1} \Bigg(\sum_{i=1}^M J_\text{f}\big(v_i(k \Delta t), T_i(k \Delta t), j_i(k \Delta t)\big) \\
		&+ \beta \bigg(J_\text{t}\Big(\big[p_1(k \Delta t), v_1(k \Delta t)\big]^\top, \big[p_\text{ref}(k \Delta t), v_\text{ref}(k \Delta t)\big]^\top\Big) \\
		 &+ \sum_{i = 2}^{M} J_\text{t}\Big(\big[p_i(k \Delta t), v_i(k \Delta t)\big]^\top, \big[\hat{p}_{i}(k \Delta t), \hat{v}_{i}(k \Delta t)\big]^\top\Big) \bigg)\Bigg),
	\end{aligned}
\end{equation}
where the weight $\beta > 0$ expresses the importance of tracking versus fuel efficiency, and $k$ is a discrete-time counter for time steps of length $\Delta t$. 
Furthermore, $\hat{p}_{i} = p_{i-1} - \zeta$ and $\hat{v}_i = v_{i-1}$ are the desired position and velocity for AV $i > 1$, with $\zeta \in \mathbb{R}$ the desired position spacing between AVs.
The tracking cost
\begin{equation}
	\begin{aligned}
		J_\text{t}(x, \hat{x}) = (x - \hat{x})^\top Q (x - \hat{x})
		\label{eq:J_t}
	\end{aligned}
\end{equation}
quadratically penalizes tracking errors, with $Q \in \mathbb{R}^{2 \times 2}$ a positive-definite weighting matrix.
The fuel cost
\begin{equation}
	J_\text{f}(v, T, j) = \Delta t \big(c^{[1]} + c^{[2]} \omega(v, j) + c^{[3]} \omega(v, j) T\big)
	\label{eq:J_f}
\end{equation}
is equal to the fuel consumption over one time step $\Delta t$.

\section{Distributed Mixed-Integer Nonlinear MPC}
\label{sec:mpc_controller}
In this section we introduce the MPC component of the proposed controller.
Define the discrete-time state and control input for AV $i$, along with the reference state, as 
\begin{equation}
	\begin{aligned}
		x_i(k) &= \big[p_i(k \Delta t), v_i(k \Delta t)\big]^\top \\
		u_i(k) &= \big[T_i(k \Delta t), F_i(k \Delta t), j_i(k \Delta t)\big]^\top \\
		x_\text{ref}(k) &= \big[p_\text{ref}(k \Delta t), v_\text{ref}(k \Delta t)\big]^\top.
	\end{aligned}
\end{equation}
Then, \eqref{eq:dynamics} can be approximated with the discrete-time dynamics $x_i(k+1) = f\big(x_i(k), u_i(k)\big)$, where
\begin{equation}
	\label{eq:disc_dynam}
	f(x, u) = \begin{bmatrix}
		x^{[1]} + \Delta t x^{[2]} \\[2pt]
		x^{[2]} + \frac{\Delta t}{m}\big(\frac{u^{[1]} z(u^{[3]}) z_\text{f}}{r} - C (x^{[2]})^2 - u^{[2]} - G\big)\end{bmatrix}.
\end{equation}

Consider a local MPC controller for the $i$th AV with prediction horizon $N > 1$ defined by the following MINLP:
\begin{subequations}
	\label{eq:MINLP}
	\begin{align}
		J\big(&x_i(k), \hat{\textbf{x}}_i(k), \textbf{p}_i^+(k), \textbf{p}_i^-(k), \textcolor{black}{j_\mathrm{prev}(k)} \big) = \nonumber \\*
		&\hspace{-12pt}\min_{\textbf{x}_i(k), \textbf{u}_i(k)} \beta \sum_{\tau = 0}^{N} J_\text{t}\big(x_i(\tau|k),\hat{x}_i(k+\tau)\big) \nonumber \\
		&\quad\quad\quad + \sum_{\tau=0}^{N-1} J_\text{f}\big(x_i^{[2]}(\tau|k), u_i^{[1]}(\tau|k), u_i^{[3]}(\tau|k)\big) \nonumber \\
		&\quad\quad\quad + \beta_\mathrm{pen} \sum_{\tau=0}^{N} \big(\sigma_i^+(\tau|k) + \sigma_i^-(\tau|k)\big) \\
		&\text{s.t.} \!\quad x_i(0|k) = x_i(k) \label{eq:IC_x}\\
		&\qquad \textcolor{black}{u^{[3]}(0|k) = j_\mathrm{prev}(k)} \label{eq:IC_j}\\
		&\qquad\text{for} \quad \tau = 0,\dots,N-1: \nonumber \\
		\vspace{0.5cm}
		&\qquad\quad x_i(\tau+1|k) = f\big(x_i(\tau|k), u_i(\tau|k)\big) \\
		&\qquad\quad |x_i^{[2]}(\tau+1|k) - x_i^{[2]}(\tau|k)| \leq a_\text{max}\Delta t \label{eq:acc_lim}\\ 
		&\qquad\text{for} \quad \tau = 0,\dots,N: \nonumber \\
		&\qquad\quad x_i^{[1]}(\tau|k) - p_i^+(\tau + k) \leq -d + \sigma_i^+(\tau|k) \label{eq:collision_ahead} \\
		&\qquad\quad x_i^{[1]}(\tau|k) - p_i^-(\tau + k) \geq d - \sigma_i^-(\tau|k) \label{eq:collision_behind} \\
		&\qquad\quad \sigma_i^+(\tau|k) \geq 0, \: \sigma_i^-(\tau|k) \geq 0 \label{eq:slacks}\\
		&\qquad\text{for} \quad \tau = 0,\dots,N-2: \nonumber \\
		&\qquad\quad |u_i^{[1]}(\tau+1|k) - u_{i}^{[1]}(\tau|k)| \leq \Delta T_\text{max} \Delta t \label{eq:T_lim} \\
		&\qquad\quad |u_i^{[3]}(\tau+1|k) - u_i^{[3]}(\tau|k)| \leq 1  \label{eq:gear_shift_lim} \\
		&\qquad\big(\textbf{x}_i(k), \textbf{u}_i(k)\big) \in \mathcal{C},
	\end{align}
\end{subequations}
where $x_i(\tau|k)$, $u_i(\tau|k)$, $\sigma_i^+(\tau|k)$, and $\sigma_i^-(\tau|k)$ are the predicted states, inputs, and slack variables for collision constraints, respectively, $\tau$ steps into the prediction window period of the MPC controller at time step $k$.
Furthermore, bold variables gather a variable over the prediction window, including \begin{equation}
	\begin{aligned}
		\textbf{x}_i(k) = \big[x_i^\top(0|k),\dots,x_i^\top(N|k)\big]^\top
	\end{aligned}
\end{equation}
and
\begin{equation}
	\begin{aligned}
		\textbf{p}_i^+ &= \big[p_i^+(k), \dots, p_i^+(k+N)\big]^\top \\
		\textbf{p}_i^- &= \big[p_i^-(k), \dots, p_i^-(k+N)\big]^\top, \\
	\end{aligned}
\end{equation}
which represent some approximation or assumption on the position of the preceding and succeeding AVs across the prediction window. 
The desired state
\begin{equation}
	\hat{\textbf{x}}_i(k) = \big[\hat{x}_i^{\top}(k),\dots,\hat{x}_i^{\top}(k+N)\big]^\top
	\label{eq:desired_state}
\end{equation}
contains the reference trajectory for the leader AV, i.e., $\hat{x}_1(\tau + k) = x_\text{ref} (\tau + k)$, and the desired spacing for all other AVs, i.e., $\hat{x}_i(\tau + k) = p_i^+(\tau + k) - \zeta$ for $i > 1$, where we recall $\zeta$ is the desired distance between an AV and its predecessor.

The constraints \eqref{eq:IC_x} and \eqref{eq:IC_j} set the initial conditions for the optimization problem, the constraint \eqref{eq:acc_lim} limits the acceleration or deceleration to $a_\text{max}$ to prevent erratic behavior, while \eqref{eq:collision_ahead} and \eqref{eq:collision_behind} maintain a safe distance of at least $d$ from the preceding and succeeding AVs.
Naturally, constraint \eqref{eq:collision_ahead} is omitted from the local MPC controller for $i = 1$, and \eqref{eq:collision_behind} omitted for $i = M$.
For feasibility, these constraints are softened with slack variables, which are penalized in the cost with a large penalty $\beta_\mathrm{pen} > 0$.
While this means that their satisfaction is not theoretically guaranteed, we note that in practice AVs can be equipped with emergency backup controllers to handle situations in which the safe distance constraints are violated \cite{nezami2021safe}; however, this is beyond the scope of the current work.
Finally, the constraint \eqref{eq:T_lim} enforces the engine torque rate constraint, and \eqref{eq:gear_shift_lim} prevents skipping gears when shifting.

The bounds on engine torque, engine speed, and brake force are grouped in
\begin{equation}
	\begin{aligned}
		\mathcal{C} = \bigg\{(\textbf{x}, \textbf{u}) \Big| &T_\text{min} \leq u^{[1]}(\tau|k) \leq T_\text{max}, \\
		&F_\text{min} \leq u^{[2]}(\tau|k) \leq F_\text{max}, \\
		&\omega_\text{min} \leq \omega\big(x^{[2]}(\tau|k), u^{[3]}(\tau|k)\big) \leq \omega_\text{max}, \\
		&\omega_\text{min} \leq \omega\big(x^{[2]}(\tau+1|k), u^{[3]}(\tau|k)\big) \leq \omega_\text{max}, \\
		&\tau = 0,\dots,N-1\bigg\}.
	\end{aligned}
\end{equation}
The last two conditions in $\mathcal{C}$, relating $x^{[2]}(\tau|k)$ and $x^{[2]}(\tau+1|k)$ to $u^{[3]}(\tau|k)$, ensures that the gear at time step $k+\tau$ maintains the engine speed within its bounds both at the beginning and at the end of each time step, i.e., both at time $(k+\tau) \Delta t$ and at time $(k+\tau+1)\Delta t$.

If no solution exists for the problem \eqref{eq:MINLP}, we set $J\big(x_i(k), \hat{\textbf{x}}_i(k), \textbf{p}_i^+(k), \textbf{p}_i^-(k), j_\mathrm{prev}(k) \big) = \infty$ by convention.

To control a platoon of AVs with $M>1$, a distributed MPC approach is considered, which involves AVs solving locally the MPC problem \eqref{eq:MINLP}, with the key consideration how to decide or agree upon the coupled variables $\hat{\textbf{x}}_i$, $\textbf{p}_i^+$, and $\textbf{p}_i^-$ \cite{mallick2024comparison}.
In platoons with no inter-AV communication, these values can be estimated based on measurements of adjacent vehicles \cite{lefevre2016learning}.
In the presence of communication, various distributed architectures can be employed to provide consistency between the true and approximate values in \eqref{eq:MINLP}.
For example, all AVs can solve \eqref{eq:MINLP} in parallel, determining $\hat{\textbf{x}}_i$, $\textbf{p}_i^+$, and $\textbf{p}_i^-$ from shifted versions of MPC solutions at the previous time step, communicated from the adjacent AVs \cite{zheng2017distributed}.
Alternatively, the AVs can solve \eqref{eq:MINLP} in a sequence, communicating the solutions down the platoon to succeeding AVs \cite{shi2017distributed}.
Finally, distributed optimization can be used where the values are agreed upon with iterations of solving and communication \cite{zhang2022semidefinite}.
We note that the methodology presented in this work is compatible with all the above distributed solution architectures\cite{mallick2024comparison}.

The MINLP \eqref{eq:MINLP} provides a state feedback controller for each AV that solves the task defined by \eqref{eq:metric}.
At time step $k $ each AV $i$ solves \eqref{eq:MINLP} numerically (potentially several times depending on the distributed paradigm) and the first element $u_i^\ast(0|k)$ of the optimal control input sequence is applied to the system.
However, the computation required to solve \eqref{eq:MINLP} numerically online renders it unsuitable for real-time implementation.
In the following, we introduce an alternative MPC controller that can be executed efficiently online.

\begin{remark}
	Note that, as \eqref{eq:MINLP} is nonconvex, in practice it is often desirable to use a multi-start approach, where the numerical optimization algorithm is run several times from different initial points, to improve the quality of the solution.
	This of course increases even further the computational burden of solving \eqref{eq:MINLP} online.
\end{remark}

\section{Distributed Parameterized Nonlinear MPC} \label{sec:parameterized_mpc}
Let us define a reduced control action that does not include the gear choice as $\mu_i(k) = \big[T_i(k \Delta t), F_i(k \Delta t)\big]^\top$.
Furthermore, define a gear-shift schedule as a vector of gears across the prediction window
\begin{equation}
	\textbf{j}_i(k) = \big[j_i(0|k),\dots,j_i(N-1|k)\big]^\top.
\end{equation}
We then introduce the following  local MPC controller for AV $i$, parameterized by $\textbf{j}_i(k)$:
\begin{subequations}
	\label{eq:NLP}
	\begin{align}
		J'&\big(x_i(k), \hat{\textbf{x}}_i(k), \textbf{p}_i^+(k), \textbf{p}_i^-(k), \textbf{j}_i(k)\big) = \nonumber \\
		&\hspace{-10pt}\min_{\substack{\textbf{x}_i(k), \bm{\mu}_i(k)}} \beta \sum_{\tau = 0}^{N} J_\text{t}\big(x_i(\tau|k),\hat{x}_i(\tau+k)\big) \nonumber \\
		&\qquad\quad + \sum_{\tau=0}^{N-1} J_\text{f}\big(x_i^{[2]}(\tau|k), \mu_i^{[1]}(\tau|k), j_i(\tau|k)\big) \nonumber \\
		&\qquad\quad + \beta_\mathrm{pen} \sum_{\tau=0}^{N} \big(\sigma_i^+(\tau|k) + \sigma_i^-(\tau|k)\big) \\
		&\;\text{s.t. } \;\;\! \eqref{eq:IC_x}, \eqref{eq:acc_lim}, \eqref{eq:collision_ahead}, \eqref{eq:collision_behind}, \eqref{eq:slacks} \\
		&\qquad\big(\textbf{x}_i(k), [\mu_i^{\top}(0|k), j_i(0|k),\dots, \nonumber \\
		&\qquad\quad\quad \mu_i^{\top}(N-1|k), j_i(N-1|k)]^\top\big) \in \mathcal{C} \label{eq:C_NLP} \\
		&\qquad\text{for } \tau = 0,\dots,N-1:\nonumber \\
		&\qquad\quad x_i(\tau+1|k) = f\big(x_i(\tau|k), [\mu_i^{\top}(\tau|k), j_i(\tau|k)]^\top\big)\label{eq:dynamics_NLP}  \\
		&\qquad\text{for } \tau = 0,\dots,N-2: \nonumber \\
		&\qquad\quad |\mu_i^{[1]}(\tau+1|k) - \mu_{i}^{[1]}(\tau|k)| \leq \Delta T_\text{max} \Delta t \label{eq:T_lim_NLP} \\
		&\qquad\quad |j(\tau+1|k) - j(\tau|k)| \leq 1 \label{eq:gear_shift_lim_NLP}.
	\end{align}
\end{subequations}
With $\textbf{j}$ prespecified, no discrete variables are optimized in the problem \eqref{eq:NLP}, which can then be solved efficiently using numerical nonlinear solvers. 
The constraint \eqref{eq:gear_shift_lim_NLP}, despite involving no decision variables, is added for consistency such that the problem \eqref{eq:NLP} is infeasible if the prespecified gear sequence $\mathbf{j}$ violates constraint \eqref{eq:gear_shift_lim}.

Note that if $\textbf{j}_i(k) = \textbf{u}_{i}^{\ast, [3]}(k)$, with $\textbf{u}_{i}^{\ast, [3]}(k)$ the optimal gear-shift schedule from \eqref{eq:MINLP}, then \eqref{eq:MINLP} and \eqref{eq:NLP} share the same set of optimal solutions.
For convenience, we denote the first element of the optimal control sequence, parameterized by $\textbf{j}_i(k)$, as $\mu_{i}^\ast\big(\textbf{j}_i(k)\big)$.

\subsection{Feasible gear-shift schedules}
We now introduce a class of gear-shift schedules for which \eqref{eq:NLP} is guaranteed to be feasible.
Define the set of feasible gears that satisfy engine speed constraints for a given velocity $v$ as:
\begin{equation}
	\Phi(v) = \Big\{j \in \{1,\dots,j_\mathrm{max}\} \Big| w_\text{min} \leq \omega(v, j) \leq w_\text{max} \Big\},
\end{equation} 
and a mapping $\phi$ which maps $v$ to one of the gears $j \in \Phi(v)$.
We highlight that, since the transmission gear ration $z$ satisfies $z(j+1) < z(j)$, given a velocity $v$ and two gears $j_1, j_2 \in \Phi(v)$ with $j_1 < j_2$, then $j\in\Phi(v), \forall j_1 \leq j \leq j_2, j\in\{1,\ldots,j_\mathrm{max}\}$.
Furthermore, define the inverse of $\Phi$, which is a set-valued mapping from gears to velocities that satisfy the engine speed constraints, as $\Omega(j) = \big\{v | j \in \Phi(v)\big\}$.
Finally, define the selection of a constant gear-shift schedule for the entire horizon $N$ via the map $\phi$ as 
\begin{equation}
	\label{eq:constant_gear_schedule}
	\rho_{\text{const}, \phi}(x) = \Big[\phi(x_i^{[2]}), \dots, \phi(x_i^{[2]})\Big]^\top.
\end{equation}
\begin{proposition}{\cite{mallick2025learning}}
	\label{prop:1}
	Assume that, for $j \in \{1,\dots,j_\mathrm{max}\}$ and for all $v \in \Omega(j)$, there exist $T$ and $F$ such that $T_\text{min} \leq T \leq T_\text{max}$, $ F_\text{max} \leq F \leq F_\text{max}$, and
	\begin{equation}\label{eq:prop1_assumption}
		\begin{aligned}
			\frac{T z(j) z_\text{f}}{r} - C v^2 - F - G = 0.
		\end{aligned}
	\end{equation}
	Then, for a state $x_i(k)$ such that $v_\mathrm{min} \leq x_i^{[2]}(k) \leq v_\mathrm{max}$, and a gear-shift sequence $\textbf{j}_i(k) = \rho_{\mathrm{const}, \phi}\big(x(k)\big)$, problem \eqref{eq:NLP} has a solution, i.e., $J'\big(x_i(k), \hat{\textbf{x}}_i(k), \textbf{p}_i^+(k), \textbf{p}_i^-(k), \textbf{j}_i(k)\big) < \infty$.
\end{proposition}
\begin{remark}
	The proof from \cite{mallick2025learning} can be applied directly here, as the MPC formulation is equivalent except for the presence of the softened constraints \eqref{eq:collision_ahead} and \eqref{eq:collision_behind}.
\end{remark}
As $(x^{[2]})^2$ for $x^{[2]} \geq 0$ is monotonic, verifying condition \eqref{eq:prop1_assumption} for a given vehicle involves checking the values of its left-hand-side only at the endpoints of the range $\Omega(j)$ for $j\in\{1,\dots,j_\mathrm{max}\}$, yielding $j_\mathrm{max}\times2$ conditions to be verified.
In general, condition \eqref{eq:prop1_assumption} is satisfied for reasonable vehicle parameters, including those used in the case study in Section \ref{sec:results}.
For these cases, Proposition \ref{prop:1} guarantees instantaneous feasibility of \eqref{eq:NLP} when considering a constant gear-shift schedule \eqref{eq:constant_gear_schedule}.
Recursive feasibility follows trivially if the true underlying system is \eqref{eq:disc_dynam}.
In the case of modeling errors, e.g., when the dynamics \eqref{eq:disc_dynam} are a discrete-time approximation of the continuous-time system \eqref{eq:dynamics}, how in the case of this paper, recursive feasibility would require a robust MPC formulation \cite{chisci2001systems}, and is left for future work.
 
Similar to \eqref{eq:MINLP}, \eqref{eq:NLP} provides a state feedback controller for each AV that solves the task defined by \eqref{eq:metric}.
At time step $k$ each AV $i$ solves \eqref{eq:NLP} (potentially several times depending on the distributed paradigm), with the first element $\mu_i^\ast(0|k)$ of the optimal control inputs and the gear $j_i(0|k)$ applied to the system.
Note that, analogous to a multi-starting approach to solve \eqref{eq:MINLP} numerically, \eqref{eq:NLP} can be solved several times in parallel for different $\textbf{j}_i(k)$, in an effort to generate a high-quality solution.
The selection of the different gear-shift schedules for which \eqref{eq:NLP} is solved is addressed next.

\section{Gear-Shift Schedule Policy}
\label{sec:policy}
Note that, for notational ease in the following, the time index $k$ is often omitted.
Proposition \ref{prop:1} guarantees that $\textbf{j}_i = \rho_{\text{const}, \phi}(x)$ is a feasible gear-shift schedule for the MPC problem \eqref{eq:NLP}.
This choice can serve as a heuristic gear-shift schedule that still allows for the co-optimization of fuel consumption and tracking via MPC.
The quality of this heuristic is then determined by the design of $\phi$, e.g., $\phi(v) = \min_{j\in\Phi(v)}j$, or $\phi(v) = \max_{j\in\Phi(v)}j$; however, in general, it will be highly suboptimal.
In light of this, we now introduce a gear-shift schedule \emph{policy} $\textbf{j}_i = \varpi_{\theta_i}(\cdot)$, defined over a suitable state-space that will be discussed in detail in the sequel, where $\varpi_{\theta_i}$ is an NN function approximator and $\theta_i$ are the model weights to be learned.
The policy $\varpi_{\theta_i}$ can be trained from data to select gear-shift schedules for \eqref{eq:NLP}. 
Once trained, $\varpi_{\theta_i}$ can then be combined with one or more heuristics $\rho_{\text{const}, \phi}$, solving \eqref{eq:NLP} for each gear-shift schedule in parallel.
The control input from the optimization problem with the lowest cost can then be applied to the system.
In this way the learned policies $\varpi_{\theta_i}$ can provide complex high-quality gear-shift schedules when the heuristic solutions become suboptimal, and the heuristic can guarantee a feasible solution to \eqref{eq:NLP} via Proposition \ref{prop:1}, something that is difficult to guarantee with learning-based policies.

\subsection{Design of \texorpdfstring{$\varpi_{\theta_i}$}{wo}}
\label{sec:design}
In this section we detail the design of the gear-shift schedule policies $\varpi_{\theta_i}$, including the input parameterization and the NN architecture.

\subsubsection{Parameterization}
\label{sec:parameterization}
An issue to overcome is that the optimal $\textbf{j}_i$ depends on the current and predicted states of the adjacent AVs $i-1$ and $i+1$, which in turn depend on their respective adjacent AVs, such that the optimal policy depends on
the states of all AVs, as well as the reference trajectory, i.e., it must have the form $\varpi_{\theta_i}(\textbf{x}_\text{ref}, x_1, \dots, x_M)$.
This is undesirable as the input dimension of the NN would then depend on the size of the platoon, and therefore the NN would require retraining with changing platoon sizes.
Furthermore, the size of the NN, and correspondingly the amount of data and training required, would grow with the size of the platoon.
Finally, every vehicle would require full knowledge of the reference trajectory and the states of all other AVs.

To address this we propose an approximate architecture that decouples $\textbf{j}_i$ from the states of the other AVs.
Define the shifted optimal solutions to \eqref{eq:NLP} at time step $k$ as 
\begin{equation}
	\begin{aligned}
		\bar{\textbf{x}}_i(k) &= \big[x_i^\top(k), x_i^{\ast, \top}(2|k-1), \dots , x_i^{\ast, \top}(N|k-1)\big]^\top \!\in \mathbb{R}^{2N} \\
		\bar{\bm{\mu}}_i(k) &= \big[\mu_i^{\ast, \top}(1|k-1),\dots, \mu_i^{\ast, \top}(N-1|k-1), \\
		&\quad\quad \mu_i^{\ast, \top}(N-1|k-1)\big]^\top \in \mathbb{R}^{2N}.
	\end{aligned}
\end{equation}
Note that the first element of $\bar{\textbf{x}}_i(k)$ is replaced with the real state $x_i(k)$, such that in the case of modeling errors the real state is present, while the last element of $\bar{\bm{\mu}}_i$ is repeated twice to ensure that $\bar{\bm{\mu}}_i$ has the same length as $\bar{\textbf{x}}_i$.
Furthermore, define the shifted gear-shift schedule
\begin{equation}
	\bar{\textbf{j}}_i(k) = \big[j_i(1|k-1), \dots, j_i(N-1|k-1), j_i(N-1|k-1)\big]^\top.
	\label{eq:shifted_gear_schedule}
\end{equation}
Consider the policy $\varpi_{\theta_i}$ to be a function of these shifted variables and the desired state as follows:
\begin{equation}
	\textbf{j}_i = \varpi_{\theta_i}(\bar{\textbf{x}}_i, \bar{\bm{\mu}}_i, \hat{\textbf{x}}_i, \bar{\textbf{j}}_i).
\end{equation}
The intuition behind this approximate parameterization is the following.
The solution of the local MPC controller \eqref{eq:NLP}, as part of a distributed MPC scheme that handles the coupled variables $\hat{\textbf{x}}_i$, $\textbf{p}_i^+$, and $\textbf{p}_i^-$, explicitly considers the coupling among AVs.
In this way, the shifted variables $\bar{\textbf{x}}_i$, $\bar{\bm{\mu}}_i, \bar{\textbf{j}}_i$ and the desired state $\hat{\textbf{x}}_i$ contain implicit information about the future behavior of the other AVs.
As a result, this parameterization, while decoupling the gear-shift schedule between AVs, is rich enough to accurately capture the optimal policy. 

\subsubsection{Architecture}
\begin{figure}
	\centering
	\includegraphics[width=0.8\columnwidth]{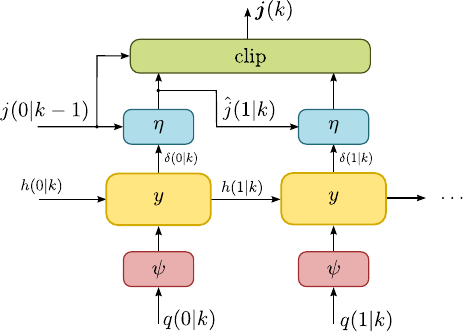}
	\caption{Recurrent NN structure. The maps $\psi$ and $\eta$ are input and output transformations, respectively.}
	\label{fig:RNN}
\end{figure}

In the following, as the policy for a single AV is described, the subscript $i$ is dropped. 
For convenience, define $q(\tau|k)$ to stack the $\tau$th elements from $\bar{\textbf{x}}(k), \bar{\bm{\mu}}(k), \hat{\textbf{x}}(k)$ and $\bar{\textbf{j}}(k)$, e.g., 
\begin{equation}
	\begin{aligned}
		&q(0|k) = \big[x^\top(k), \mu^{\ast, \top}(1|k-1), \hat{x}^\top(k), j(1|k-1)\big]^\top, \\
		&\qquad\qquad\qquad\qquad\qquad\qquad\cdots \\
		&q(N-1|k) = \big[x^{\ast, \top}(N|k-1) , \mu^{\ast, \top}(N-1|k-1), \\
		&\quad\quad\quad\quad\quad\quad \hat{x}^\top(k+N-1), j(N-1|k-1)\big]^\top.
	\end{aligned}
\end{equation}
For $\varpi_{\theta}$ we propose an NN where, to limit the action space and force incremental gear shifts, the outputs represent a sequence of down-shift, no-shift, and up-shift commands, which are then mapped to a gear-shift schedule. 
To this end, using a standard feed-forward NN has the key issue that the action space grows exponentially with the prediction horizon $N$, while the input space grows linearly.
Indeed, for a given $N$ there are $3^N$ possible sequences of up/down/no-shift commands, and $7N$ inputs (from $\bar{\textbf{x}},\hat{\textbf{x}}, \bar{\bm{\mu}},$ and $\bar{\textbf{j}}$).
An NN capable of representing this input-output mapping as $N$ increases may need to be very large and highly complex.
Furthermore, there is an explicit temporal relationship between gear-shifts, which is not structurally enforced in a feed-forward NN.
Recurrent NNs (RNNs) can be used to overcome this form of explosion in the complexity of the input-output mapping \cite{mallick2025learning, da2024integrating}. 
Therefore, here we propose to adopt a sequence-to-sequence recurrent architecture using an RNN, as shown in Figure \ref{fig:RNN}.
The elements of the inputs $\bar{\textbf{x}}, \bar{\bm{\mu}}, \hat{\textbf{x}}$, and $\bar{\textbf{j}}$ are considered as $N$ different inputs in a chain consisting of the vectors $q(\tau|k) \in \mathbb{R}^6\times\{1,\ldots,j_\mathrm{max}\}$ for $\tau=0,\dots,N-1$.
A single RNN is trained with input space $\mathbb{R}^6\times\{1,\ldots,j_\mathrm{max}\}$, where the output is a single shift command $\{-1, 0, 1\}$.
The output sequence of $N$ gear positions is then generated by sequentially evaluating the RNN on the chain of inputs $q(0|k),\dots, q(N-1|k)$.
In this way the recurrent structure results in a constant number of inputs and outputs for any prediction horizon, with only the number of sequential evaluations changing as the horizon changes.
Furthermore, the temporal relationship is structurally enforced, i.e., the gear at time step $\tau+k$ considers the prior gears and inputs via the hidden state $h_\tau$.

More formally, the policy is defined as
\begin{equation}
	\begin{aligned}
		&\varpi_{\theta}\big(\bar{\textbf{x}}(k), \bar{\bm{\mu}}(k), \hat{\textbf{x}}(k), \bar{\textbf{j}}(k)\big) = \Bigg[\eta\bigg(y_\theta\Big(\psi\big(q(0|k)\big), h_0\Big)\bigg), \\
		&\quad  \dots,\eta\bigg(y_\theta\Big(\psi\big(q(N-1|k)\big), h_{N-1}\Big)\bigg)\Bigg]^\top,
	\end{aligned}
\end{equation}
where the input mapping $\psi$, defined by
\begin{equation}
	\begin{aligned}
		\psi\big(q = [x^\top, \mu^\top, \hat{x}^{\top}, &j]^\top\big) = \bigg((x-\hat{x})^\top, \frac{x^{[2]} - v_\text{min}}{v_\text{max} - v_\text{min}}, \\
		&\quad\frac{\hat{x}^{[2]} - v_\text{min}}{v_\text{max} - v_\text{min}}, \mu^\top, \omega(x^{[2]}, j), j\bigg)^\top,
	\end{aligned}
\end{equation}
transforms the inputs into a representation that contains the tracking error, the vehicle and target normalized velocities, the predicted inputs, and the predicted engine speed.
This representation is chosen to give the RNN the most relevant information for selecting a gear-shift schedule.
The function 
\begin{equation}
	\delta(\tau|k) = \big[\delta_\text{d}(\tau|k), \delta_\text{n}(\tau|k), \delta_\text{u}(\tau|k)\big]^\top = y_\theta\Big(\psi\big(q(\tau|k)\big), h_\tau\Big)
\end{equation} is the model function of the RNN, where $\delta_\text{d/n/u}(\tau|k)$ is a score related to choosing down/no/up-shift at the $\tau$th output in the sequence.
The output mapping $\eta$ applies the shift command with the largest score $\delta$ to the gear $\hat{j}$ that was selected at the previous time step along the prediction window, i.e.,
\begin{equation}
	\eta\big(\delta(\tau|k), \hat{j}(\tau|k)\big) = \hat{j}(\tau|k) + \varphi\big(\delta(\tau|k)\big),
	\label{eq:policy_output_mapping}
\end{equation}
where $\hat{j}(0|k) = j(0|k-1)$, $\hat{j}(\tau|k) = \eta\big(\delta(\tau-1|k), \hat{j}(\tau-1|k)\big)$ for $\tau > 0$, and $\varphi$ is defined\footnote{We remark that the $-2$ in \eqref{eq:gear_selection_from_probabilities} translates the index of the highest score in the vector $\delta$ to the corresponding gear-shift action in the set $\{-1, 0, 1\}$.} as
\begin{equation}
	\varphi\big(\delta(\tau|k)\big) = \argmax_{i\in\{1, 2, 3\}} \delta^{[i]}(\tau|k) - 2.
	\label{eq:gear_selection_from_probabilities}
\end{equation}
Finally, clipping is applied to force the gears within the range $\{1,\ldots,j_\mathrm{max}\}$:
\begin{equation}
	j(\tau|k) = \text{clip}\Big(\eta\big(\delta(\tau|k), \hat{j}(\tau)\big), 1, j_\mathrm{max}\Big).
\end{equation}

\subsection{Training of \texorpdfstring{$\varpi_{\theta_i}$}{wo}}
\label{sec:training}
\subsubsection{Supervised versus reinforcement learning}
In general, $\varpi_{\theta_i}$ can be trained in an SL or RL fashion.
Training with SL requires input-output data of the form $\bar{\textbf{x}}_i, \bar{\bm{\mu}}_i, \hat{\textbf{x}}_i, \bar{\textbf{j}}_i \to \textbf{j}_i^\ast$, the generation of which involves evaluating the distributed MPC scheme using the MINLP \eqref{eq:MINLP} to determine $\textbf{j}_i^\ast$.
The computational complexity of \eqref{eq:MINLP} renders this undesirable, particularly as the size of the platoon grows.
Furthermore, SL is limited to training $\varpi_{\theta_i}$ to approximate the solution to \eqref{eq:MINLP}, and is limited by the accuracy of the model \eqref{eq:disc_dynam} used in \eqref{eq:MINLP}.

In contrast, we propose to train $\varpi_{\theta_i}$ with RL, in which a policy is trained with data generated from interactions with the real (or simulated) system, avoiding solving \eqref{eq:MINLP}.
Furthermore, an RL approach facilitates tailoring the learning task via crafting the reward signal, allowing us to go beyond simply approximating the solution to \eqref{eq:MINLP}.
This is useful for embedding certain complexities in the policy, e.g., providing additional reward when $\varpi_{\theta_i}$ gives a better solution than $\rho_{\text{const}, \phi}$ encourages this behavior.
Finally, as the data is generated via interaction with the system, the resulting policy is not entirely dependent on the quality of the MPC model. 

\subsubsection{From multi-agent RL to single-agent RL}
At first glance, training $\varpi_{\theta_i}$ for $i = 1,\dots,M$ with RL results in a MARL problem.
In this context, as AVs learn while interacting with other learning AVs, the learning task becomes a non-stationary MARL problem \cite{busoniu2008comprehensive}.
In general this significantly increases the learning challenge, with extensive research dedicated to develop algorithms that mitigate the issue \cite{lowe2017multi, yu2022surprising}.
Equally challenging, data collection for a platoon may be less accessible than for a single AV. 

To avoid a MARL solution, we propose to train $\varpi_{\theta_i}$ in a single-vehicle-following scenario. 
Indeed, the decoupled parameterization detailed in Section \ref{sec:parameterization} naturally allows for the policies to be trained independently, as the RNN is a function of only locally available information.
Consider an independent single-AV MPC controller
\begin{subequations}
	\label{eq:NLP_single}
	\begin{align}
		\tilde{J}\big(x_i(k), &\textbf{x}_\text{ref}(k), \textbf{j}_i(k)\big) = \nonumber \\
		&\hspace{-12pt}\min_{\substack{\textbf{x}_i(k), \bm{\mu}_i(k)}} \beta \sum_{\tau = 0}^{N} J_\text{t}\big(x_i(\tau|k),\hat{x}_i(\tau+k)\big) \nonumber \\
		&\quad + \sum_{\tau=0}^{N-1} J_\text{f}\big(x^{[2]}_i(\tau|k), \mu_i^{[1]}(\tau|k), j_i(\tau|k)\big) \\
		&\text{s.t.} \quad \eqref{eq:IC_x}, \eqref{eq:acc_lim}, \eqref{eq:C_NLP}, \eqref{eq:dynamics_NLP}, \eqref{eq:T_lim_NLP}	\end{align}
\end{subequations}
with the first element of the optimal control sequence, parameterized by $\textbf{j}_i(k)$, again denoted with $\mu_i^\ast\big(\textbf{j}_i(k)\big)$ for convenience.
We propose to train $\varpi_{\theta_i}$ in a scenario where the AV tracks a reference trajectory using the MPC controller \eqref{eq:NLP_single} in the closed loop.
We argue that a policy trained in this way can then generalize to the platoon case, as the relevant information pertaining to the other AVs is embedded in the shifted variables that serve as inputs to the policy.
Note that, if the AVs are homogeneous, one policy can be trained and used for each AV in the platoon in a parameter-sharing fashion \cite{tan1993multi}. 
In the non-homogeneous case, each AV must train a separate policy, but it can do so in the above single-vehicle-following scenario.

\subsubsection{Deep Q-Learning}
\begin{figure*}[t]
	\centering
	\includegraphics[width=0.7\textwidth]{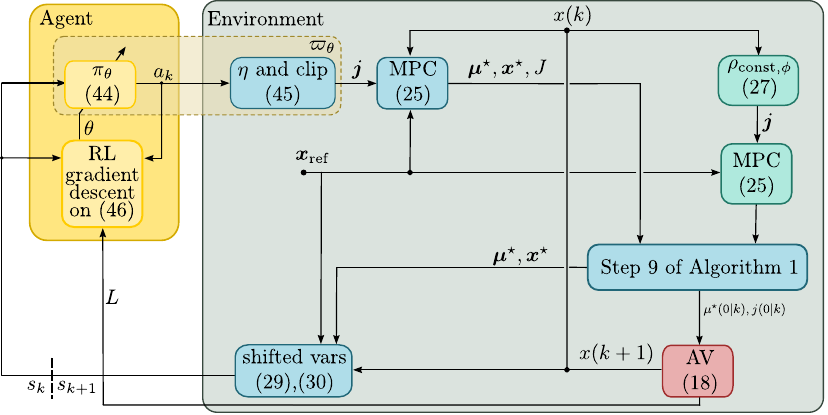}
	\caption{RL training setup. At the end of the training process, the policy that gets deployed is $\varpi_\theta$, indicated by the dashed box.}
	\label{fig:train}
\end{figure*}

Let us formalize the RL problem for learning $\varpi_{\theta_i}$.
For simplicity, in the following we again drop the subscript $i$.
Consider a Markov Decision Process (MPD) \cite{sutton2018reinforcement} with state
\begin{equation}
	s = (\bar{\textbf{x}}, \bar{\bm{\mu}}, \textbf{x}_\text{ref}, \bar{\textbf{j}}) \in \mathcal{S} = \mathbb{R}^{2N} \times \mathbb{R}^{2N} \times \mathbb{R}^{2N} \times \{1,\dots,j_\mathrm{max}\}^N
\end{equation}
and discrete action representing a down/no/up-shift sequence
\begin{equation}
	a \in \mathcal{A} = \{-1, 0, 1\}^N.
	\label{eq:rl_action_def}
\end{equation}
The MDP has deterministic state transitions modeled by the state transition function $s_+ = F(s, a)$ which absorbs both the MPC controller and the AV dynamics.
Specifically, the action $a$ generates a gear-shift schedule $\mathbf{j}$, the NLP \eqref{eq:NLP_single} is solved, the control input $[\mu^*(\mathbf{j})^\top, j(0)]^\top$ is applied to the system, and the solutions to the optimization problem are shifted to form the next state $s_+$, as depicted in Figure \ref{fig:train}.

We consider a deterministic policy $\pi_\theta : \mathcal{S} \to \mathcal{A}$ that maps a state of the MPD to an action, i.e., a shift command sequence.\footnote{We highlight the distinction between the RL policy $\pi$, which selects a sequence $a$ of down/no/up-shift commands, and the gear-shift schedule policy $\varpi$.
The connection between the two is made explicit in the sequel.}
We propose to use the value-based method Deep Q-Network (DQN) \cite{mnih2015human}, as it is well suited for discrete action spaces.
However, in light of the proposed RNN architecture, DQN cannot be applied directly to learn the policy.
In fact, the outputs of the RNN defined in Section \ref{sec:design} and those of $Q_\pi(s, a)$ do not match. 
The output of the RNN is a single value associated to a sub-action in the sequence, i.e., a singular down/no/up-shift command, rather than a value for the entire action (the shift command \emph{sequence} $a$ introduced in \eqref{eq:rl_action_def}).
Indeed, analogous to the scaling issue highlighted in Section \ref{sec:design}, a function approximator for $Q$ that outputs a value for each possible $a \in \mathcal{A}$ requires a number of outputs that is exponential in the horizon $N$, and is in general intractable.
In light of this, we take inspiration from \cite{da2024integrating} and shift the goal to learning a proxy for decoupled Q-values, i.e., the value associated with each sub-action $\tilde{a}\in\{-1,0,1\}$ in the sequence $a_k=\big[\tilde{a}(0|k),\ldots,\tilde{a}(N-1|k)\big]^\top$, denoted by
\begin{eqnarray}
	\tilde{Q}_\theta\big(q(\tau|k), \tilde{a}(\tau|k), h_\tau\big), 
\end{eqnarray}
where $q(\tau|k)$, depicted in Figure \ref{fig:RNN}, is the $\tau$th element in the input sequence.
As in \cite{da2024integrating}, we argue that this can accurately approximate the DQN algorithm applied to learn the full action-value function, as the hidden states $h_\tau$ contain information pertaining to the inputs and outputs at the other stages of the RNN sequence.
Note then that the structure for $\varpi_\theta$ proposed in Section \ref{sec:design} naturally fits this learning goal, with $\Tilde{Q}_\theta$ and $\pi_\theta$ defined explicitly as
\begin{equation}
	\tilde{Q}_\theta(q, \tilde{a}, h) = \bigg(y_\theta\big(\psi(q), h\big)\bigg)^{[\tilde{a}+2]}
\end{equation}
and 
\begin{equation}
	\label{eq:policy}
	\begin{aligned}
		\pi_\theta(s_k) &=\bigg[\varphi\Big(y_\theta\big(\psi(q(0|k)), h_0\big)\Big), \dots, \\
		&\quad\quad\varphi\Big(y_\theta\big(\psi(q(N-1|k)), h_{N-1}\big)\Big)\bigg]^\top.
	\end{aligned}
\end{equation}
The gear-shift schedule policy can then be expressed equivalently as:
\begin{equation}
	\label{eq:gear_shift_policy}
	\begin{aligned}
		\varpi_\theta(s_k) &= \text{clip}\Bigg(\bigg[\big(\pi_\theta(s_k)\big)^{[1]} + j(0|k-1), \dots, \\ &\quad\sum_{\tau=0}^{N-1}\Big(\big(\pi_\theta(s_k)\big)^{[\tau]}\Big) + j(0|k-1)\bigg]^\top, 1, j_\mathrm{max}\Bigg),
	\end{aligned}
\end{equation}
where the summation originates from the iterative application of \eqref{eq:policy_output_mapping} along the prediction window.

The DQN algorithm is then applied as described in Section \ref{sec:background}.
Given the use of $\Tilde{Q}_\theta$ in place of  $Q_\theta$, the loss function \eqref{eq:loss} is modified as:
\begin{equation}
	\label{eq:loss_q_tilde}
	\begin{aligned}
		L_{\text{loss}}(\theta) = & \sum_{l=1}^{N_\text{batch}} \sum_{\tau=0}^{N-1} \ell \bigg(L(s^{(l)}_k, a^{(l)}_k) \\
		&+ \gamma \max_{\hat{a}\in\{-1, 0, 1\}} \tilde{Q}_{\tilde{\theta}}\big(q^{(l)}(\tau|k+1), \hat{a}, h_\tau\big) \\
		&- \tilde{Q}_\theta\big(q^{(l)}(\tau|k), \tilde{a}^{(l)}(\tau|k), h_\tau\big)\bigg).
	\end{aligned}
\end{equation}

\subsubsection{RL cost function}
\label{sec:two_stage_train}

Note that the exact form of the learning task depends on the form of the stage cost $L$.
Indeed, designing $L$ such that the resulting policy performs well in terms of the performance \eqref{eq:metric} is a design challenge.
We propose the general form 
\begin{equation}\label{eq:RL_cost}
	\begin{aligned}
		L(s_k, a_k) &= \beta J_\text{t}\big(x(k), \textbf{x}_\text{ref}(k)\big) \\
		 &\quad + J_\text{f}\Big(x^{[2]}(k), \mu^{\ast, [1]}\big(\varpi_\theta(s_k)\big), \varpi^{[1]}_\theta(s_k)\Big) - e \cdot \kappa,
	\end{aligned}
\end{equation}
penalizing the tracking error and fuel consumption.
The optional indicator $\kappa$ allows the shaping of the cost to encourage certain behaviors, e.g., penalizing infeasible gear-shift sequences, with $e$ a weight.
An example of how $\kappa$ can be designed to improve the quality of the resulting policy is found in Section \ref{sec:results}.

\subsection{Summary of Training Procedure}
In Algorithm \ref{alg:training_1} we summarize the training, while Algorithm \ref{alg:controller} summarizes how the policy is used at deployment in a platoon, from the perspective of a single AV.
In steps \ref{eq:step_9} to \ref{eq:step_13} in Algorithm \ref{alg:training_1} the heuristic solution $\rho_{\mathrm{const}, \phi}\big(x(k)\big)$ is used in the case of infeasibility of $\varpi_\theta(s_k)$.
Conversely, in steps \ref{eq:step_5} to \ref{eq:step_9_alg_2} of Algorithm \ref{alg:controller} the heuristic solution $\rho_{\mathrm{const}, \phi}\big(x(k)\big)$ is used whenever it outperforms $\varpi_\theta(s_k)$.
Finally, in steps \ref{step:dmpc_1} and \ref{step:dmpc_2} of Algorithm \ref{alg:controller} `distributed MPC scheme' is a placeholder for any distributed MPC methodology that coordinates the coupled variables of the AVs, e.g., parallel \cite{zheng2017distributed}, serial \cite{shi2017distributed}, or iterative \cite{zhang2022semidefinite} methodologies.
See Figure \ref{fig:train} for a schematic depiction of the training process.
In the next section, the specific methodology used in this paper's case study is discussed.

\begin{algorithm}
	\caption{Training at time step $k$}
	\label{alg:training_1}
	\begin{algorithmic}[1]
		\State \textbf{Inputs}: Current state $x(k)$, shifted solutions $\bar{\textbf{x}}(k)$, $\bar{\bm{\mu}}(k)$, $\bar{\textbf{j}}(k)$, transition buffer $\mathcal{D}$, exploration probability $\epsilon(k)$, reference trajectory $\textbf{x}_\text{ref}(k)$
		\State $s_k \gets \big(\bar{\textbf{x}}(k), \bar{\bm{\mu}}(k), \textbf{x}_\text{ref}(k), \bar{\textbf{j}}(k)\big)$ \label{eq:step_2}
		\If{$\varepsilon > \epsilon(k)$ with $\varepsilon \sim \mathcal{U}(0, 1)$}
			\State $a_k \gets \pi_\theta(s_k)$ as in \eqref{eq:policy}
		\Else
			\State $a_k \sim \mathcal{U}\big(\{-1, 0, 1\}^{N}\big)$
		\EndIf 
		\State Form $\varpi_\theta(s_k)$ from $a_k$ as in \eqref{eq:gear_shift_policy}\label{eq:step_7}
		\If{$\tilde{J}\big(x(k), \textbf{x}_\text{ref}(k), \varpi_\theta(s_k)\big)  < \infty$}\label{eq:step_9}
			\State $u \gets \big[\mu^{\ast, \top}\big(\varpi_\theta(s_k)\big), \varpi_\theta^{[1]}(s_k)\big]^\top$
		\Else
			\State $u \gets \big[\mu^{\ast, \top}\big(\rho_{\text{const}, \phi}\big(x(k)\big)\big), \rho_{\text{const}, \phi}^{[1]}\big(x(k)\big)\big]^\top$
		\EndIf \label{eq:step_13}
		\State Compute $\kappa$
		\State Apply $u$ to the system and observe $x(k+1)$, $J_\text{t}\big(x(k), x_\text{ref}(k)\big)$ and $J_\text{f}\big(x^{[2]}(k), u^{[1]}, u^{[3]}\big)$ \label{eq:step_15}
		\State Form shifted solutions $\bar{\textbf{x}}(k+1)$, $\bar{\bm{\mu}}(k+1),\bar{\textbf{j}}(k+1)$
		\State $s_{k+1} \gets \big(\bar{\textbf{x}}(k+1), \bar{\bm{\mu}}(k+1), \textbf{x}_\text{ref}(k+1), \bar{\textbf{j}}(k+1)\big)$ \label{eq:step_17}
		\State $L(s_k, a_k) \gets \beta J_\text{t}\big(x(k), x_\text{ref}(k)\big) + J_\text{f}\big(x^{[2]}(k), u^{[1]}, u^{[3]}\big) + e \cdot \kappa$
		\State Store transition $\big(s_k, a_k, L(s_k, a_k), s_{k+1}\big)$ in $\mathcal{D}$\label{eq:step_19}
		\State Sample $N_\text{batch}$ samples from $\mathcal{D}$ and perform a gradient descent step on \eqref{eq:loss_q_tilde} to update $\theta$
		\State Update the target network weights \eqref{eq:target_blending} \label{eq:step_22}
	\end{algorithmic}
\end{algorithm}
\begin{algorithm}
	\caption{Controller at time step $k$ for AV $i$}
	\label{alg:controller}
	\begin{algorithmic}[1]
		\State \textbf{Inputs}: Current state $x_i(k)$, shifted solutions $\bar{\textbf{x}}_i(k)$, $\bar{\bm{\mu}}_i(k)$, $\bar{\textbf{j}}_i(k)$, desired state $\hat{\textbf{x}}_i(k)$
		\State $s_{k,i} \gets \big(\bar{\textbf{x}}_i(k), \bar{\bm{\mu}}_i(k), \hat{\textbf{x}}_i(k), \bar{\textbf{j}}_i(k)\big)$ 
		\State Apply distributed MPC scheme to solve \eqref{eq:NLP} and determine $J'_1  \gets J'\big(x_i(k), \hat{\textbf{x}}_i(k), \textbf{p}_i^+(k), \textbf{p}_i^-(k), \varpi_{\theta_i}(s_{k, i})\big)$ \label{step:dmpc_1}
		\State Apply distributed MPC scheme to solve \eqref{eq:NLP} and determine $J'_2  \gets J'\Big(x_i(k), \hat{\textbf{x}}_i(k),\textbf{p}_i^+(k), \textbf{p}_i^-(k), \rho_{\text{const}, \phi}\big(x(k)\big)\Big)$ \label{step:dmpc_2}
		\If{$J'_1  < J'_2$} \label{eq:step_5}
		\State $u_i \gets \Big[\mu_i^{\ast, \top}\big(\varpi_{\theta_i}(s_{k,i})\big), \varpi_{\theta_i}^{[1]}(s_{k,i})\Big]^\top$
		\Else
		\State $u_i \gets \Big[\mu_i^{\ast, \top}\big(\varpi_{\theta_i}(s_{k,i})\big), \rho_{\text{const}, \phi}^{[1]}\big(x_i(k)\big)\Big]^\top$
		\EndIf \label{eq:step_9_alg_2}
		\State Apply $u_i$ to the system
	\end{algorithmic}
\end{algorithm}

\section{Experiments}
\label{sec:results}
We illustrate the ideas presented in this work with a sequential distributed approach \cite{mallick2024comparison, shi2017distributed}.
Specifically, in a sequential approach the AVs solve the local MPC problems in succession, from the front of the platoon to the end.
For AV $i$ at time step $k$, the value for $\textbf{p}_i^+(k)$ is determined from $\textbf{x}_{i-1}^{\ast}(k)$, communicated from AV $i-1$ at time step $k$, as:
\begin{equation}
	\begin{aligned}
		\textbf{p}_i^+(k) = \big[x_{i-1}^{[1], \ast}(0|k), \dots, x_{i-1}^{[1], \ast}(N|k)\big]^\top.
	\end{aligned}
\end{equation}
Similarly, the value for $\textbf{p}_i^-(k)$ determined from $\bar{\textbf{x}}_{i+1}(k)$, communicated from AV $i+1$ at time step $k-1$, as:
\begin{equation}
	\textbf{p}_i^-(k) = \big[x_{i+1}^\top(k), x_{i+1}^{\ast, \top}(2|k-1), \dots , x_{i+1}^{\ast, \top}(N|k-1)\big]^\top.
\end{equation}
We note again that the methodology presented in this paper can be combined with any distributed solution architecture.

\subsection{Comparison Controllers}
In the following we describe the form of the local controllers that are solved sequentially for each control method used in the comparison.
For more details, see the appendix of the extended online version of this article.
\begin{itemize}
	\item \textbf{MINLP-based MPC (MINLP)}: This local MPC controller solves the MINLP \eqref{eq:MINLP} at each time step $k$, applying $u_i^\ast(0|k)$ to the system.
	\item \textbf{MIQP-based MPC (MIQP) \cite{shao2021vehicle}}: This local controller applies convexifications to the fuel model and vehicle dynamics such that the resulting optimization problem is an MIQP.
	Convexification introduces errors; however, in general it lowers the computational burden.
	\item \textbf{Heuristic Decoupled MPC (HD)}: This local controller follows the principle of decoupling the optimization of the vehicle speed from the gear-shift schedule.
	A simplified dynamic model is used within the MPC controller, optimizing only the continuous vehicle dynamics, as fuel consumption is unmodeled, with the gear selected solely based on the resulting vehicle velocity.
	\item \textbf{Heuristic Co-optimization MPC (HC)}: This local controller solves the NLP \eqref{eq:NLP} for gear-shift schedules selected by three choices of the heuristic $\textbf{j}_i(k) = \rho_{\text{const}, \phi_i}\big(x_i(k)\big)$ for $i = 1,2,3$, with $\phi_1, \phi_2$, and $\phi_3$ corresponding to taking the lowest, highest, and middle of the feasible gear choices.
	\item \textbf{Heuristic Shifted-solution Co-optimization MPC (HS)}: This local controller solves the NLP \eqref{eq:NLP} with the gear-shift schedule obtained by shifting the gear-shift schedule from the previous time step and selecting the last element via $\phi_2$.
	\item \textbf{Learning Co-optimization MPC (LC)}: This local controller solves the NLP \eqref{eq:NLP} in parallel for the three heuristic gear-shift schedules $\phi_1, \phi_2, \phi_3$ and for $\textbf{j}_i(k) = \varpi_{\theta_i}(s_{k, i})$, where $\varpi_{\theta_i}$ has been trained with Algorithm \ref{alg:training_1}.
\end{itemize}

\subsection{Experiment Details}
The training of $\varpi_\theta$ and the controller evaluations have been implemented in Python 3.11.13 and run on a Linux server with 8 AMD EPYC 7252 (3.1 GHz) processors, four Nvidia GeForce RTX 3090 GPUs, and 251 GB of RAM.
MINLP problems are solved with Knitro \cite{byrd2006knitro}, MIQP problems are solved with Gurobi \cite{gurobi}, and NLP problems are solved with Ipopt \cite{wachter2006implementation}.
These solvers are each state-of-the-art for the respective type of optimization problem.
Source code is available at \cite{github-repo}.

We consider a platoon of homogeneous AVs defined by the coefficients and variable bounds given in Table \ref{tab:dynamics_coefficients}.
\begin{table}
	\centering
	\caption{Constants/variables for the vehicle dynamics \cite{shao2021vehicle}.}
	\label{tab:dynamics_coefficients}
	\begin{tabular}{ccc}
		\hline
		\textbf{Symbol} & \textbf{Value/Bounds} & \textbf{Unit} \\ 
		\hline
		$m$ & $2000$ & kg \\
		$C$ & $0.4071$ & kg/m\\
		$\mu$ & $0.015$ & - \\
		$g$ & $9.81$ & m/s$^2$\\
		$z_\text{f}$ & $3.39$ & - \\
		$r$ & $0.3554$ & m \\
		$j_\mathrm{max}$ & 6 & - \\
		$\Delta T_\text{max}$ & $100$ & Nm/s \\
		$c$ & $[0.04981, 0.001897, 4.5232 \cdot 10^{-5}]^\top$ & see \cite{shao2021vehicle} \\
		$a$ & $[-3, 3]$ & m/s$^2$ \\
		$v$ & $[2.204, 44.388]$ & m/s \\
		$F_\text{b}$ & $[0, 9000]$ & N \\
		$T$ & $[15, 300]$ & Nm \\
		$w$ & $[900, 3000]$ & RPM \\
		$z$ & $[4.484, 2.872, 1.842, 1.414, 1, 0.742]$ & - \\
		$m_\text{f}$ & - & L \\
		\hline
	\end{tabular}
\end{table}
The platoon operates in a high-way driving scenario where the leader AV follows a reference trajectory that is randomly generated with the acceleration changing at each time step with probability $1/20$ and taking values uniformly sampled from the range $[-3, 3]$ m/s$^{2}$.
Furthermore, the velocity is clipped to the range $[5, 28]$ m/s (18-100 km/h).
Each of the vehicles in the platoon, except for the leader, follows the one in front, with a desired position spacing of $\zeta = 25$ m. The safety distance $d$ is set to 10 m. 
For all simulations we have selected $\beta=0.01$, tuned to balance the relative importance of the fuel consumption and the quadratic tracking error, with $Q=\text{diag}(1, 0.1)$.
The slack variable penalty is selected as $\beta_\text{pen}=1000$.
All the MPC controllers operate with a time step of $\Delta t = 1$s.
We recall that a sequential distributed control architecture is used, where the AVs solve their local control problems in a sequence, from the leader to the last AV, each communicating their solution to the succeeding AV \cite{shi2017distributed}. 

As nonlinear programs are solved, multi-start strategies can improve solution quality.
We select the number of multi-start points for each controller in order to give a fair comparison.
The HC and LC approaches inherently include the notion of multi-starting by solving the optimization problem in parallel for different gear schedules, therefore for these controllers we opt not to use another level of multi-starting, such that these controllers solve 3, 4, and 4 NLPs in parallel, respectively.
For the HD and HS controllers, 4 multi-start points are used such that, again, 4 NLPs are solved in parallel.
For the MIQP approach, as a convex-mixed integer program with a unique optimum for each configuration of integer variables is solved, no multi-starting is used.
For the MINLP controller, we again use 4 multi-start points. 
Note that all controllers one optimization problem is warm started using a shifted version of the solution from the previous time step, with additional multi-start-points selected randomly.

The reported solve times in the following refer to the time required to decide a control action of the whole platoon, which, with the sequential distributed MPC methodology, is the sum of the solve times for each local MPC problem.
Furthermore, due to the extreme computational burden of solving the MINLP problem, the wall-time, with multi-start solution computed in parallel, is limited to 600s per vehicle.

\subsection{Training}
For the gear-shift schedule policy we use an RNN with 4 layers of 256 features in the hidden state, followed by a fully connected linear layer.
During training a single AV tracks the randomly generated reference trajectory where, if the position tracking error exceeds $100$m, the reference trajectory is reset to the AVs current state, thus avoiding exploding RL costs during the early phases of training.

For training, we adopt a two-stage approach that emphasizes different learning objectives via the indicator $\kappa$ in the RL cost.
It is reasonable to assume that the policy $\varpi_\theta$, initialized randomly, is very poor, and is likely to suggest infeasible gear-shift schedules frequently.
Therefore, in the first stage the emphasis is placed on learning to select feasible gear-shift sequences.
To this end an indicator $\kappa_1$ is defined as
\begin{equation}
	\kappa_1 = \begin{cases}
		0 & \text{if } \tilde{J}\big(x, \textbf{x}_\text{ref}, \varpi_\theta(s)\big)  < \infty \\ 
		1 & \text{if } \tilde{J}\big(x, \textbf{x}_\text{ref}, \varpi_\theta(s)\big)  = \infty
	\end{cases}.
	\label{eq:kappa_stage_1}
\end{equation}
For the second stage, the goal is to refine the policy's behavior under deployment conditions, i.e., when the best gear-shift schedule between the RL policy and the heuristic is used, as in steps \ref{eq:step_5} to \ref{eq:step_9_alg_2} of Algorithm \ref{alg:controller}.
To this end an indicator $\kappa_2$ is defined as
\begin{equation}
	\kappa_2 = \begin{cases}
		1 & \text{if } \tilde{J}\big(x, \textbf{x}_\text{ref}, \varpi_\theta(s)\big)  \leq \tilde{J}\big(x, \textbf{x}_\text{ref}, \rho_{\text{const}, \phi}(x)\big)  \\ 
		0 & \text{if } \tilde{J}\big(x, \textbf{x}_\text{ref}, \varpi_\theta(s)\big) > \tilde{J}\big(x, \textbf{x}_\text{ref}, \rho_{\text{const}, \phi}(x)\big) 
	\end{cases},
	\label{eq:kappa_stage_2}
\end{equation}
with the input applied being that of the best solution:
\begin{equation}
	u = \begin{cases}
		\big[\mu^{\ast, \top}\big(\varpi_\theta(s)\big), \varpi_\theta^{[1]}(s)\big]^\top & \text{if } \kappa_2 = 1 \\ 
		\big[\mu^{\ast, \top}\big(\rho_{\text{const}, \phi}(x)\big), \rho_{\text{const}, \phi}^{[1]}(x)\big]^\top & \text{if } \kappa_2 = 0
	\end{cases}.
	\label{eq:u_stage_2}
\end{equation}
This stage encourages the policy to outperform the heuristic, and infeasibility is not punished.
In summary, stage 1 has the goal of learning a policy that is feasible wherever possible, while stage 2 has the goal of refining the policy and improving it by competing with the heuristic.
We refer to the RL cost \eqref{eq:RL_cost} for stage 1 as $L_1$, with $e_1$ the weight associated to $\kappa_1$, while $L_2$ and $e_2$ are the corresponding variables for stage 2.
Table \ref{tab:training_coefficients} gives the hyperparameters used for training $\varpi_\theta$ using Algorithm \ref{alg:training_1}
\begin{table}[ht]
	\centering
	\caption{Hyperparameters used for training.}
	\label{tab:training_coefficients}
	\begin{tabular}{cc|cc}
		\hline
		\textbf{Parameter} & \textbf{Value} & \textbf{Parameter} & \textbf{Value} \\ 
		\hline
		$\gamma$ & $0.9$ & 	$e_1$ & $10^4$\\
		$\alpha$ & $0.001$ & $e_2$ & $100$\\
		$\nu$ & $0.001$ & $\mathcal{D}$ max size & $100000$\\
		$\epsilon(k)$ & $0.99e^{-2.76\cdot 10^{-6}k}$ & $N_\text{batch}$ & 128 \\
		\hline
	\end{tabular}
\end{table}

\begin{figure}[t]
	\centering
	\input{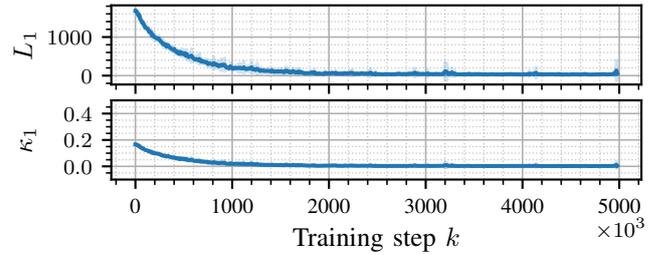}
	\vspace{-0.6cm}
	\caption{Performance metrics over the training stage 1. The plots show mean (solid) and standard deviation (shaded) of the moving average of the metrics for a moving window of 10000 steps.}
	\label{fig:train_stage_1}
\end{figure}

\begin{figure}[t]
	\centering
	\input{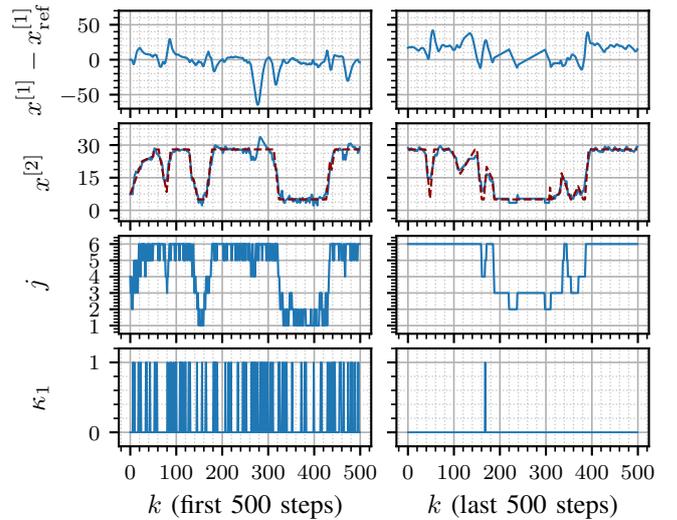}
	\vspace{-0.6cm}
	\caption{Representative trajectories during training stage 1. The red dashed line in the second row represents $x^{[2]}_\mathrm{ref}$.}
	\label{fig:training_traj}
\end{figure}

Figure \ref{fig:train_stage_1} shows the costs $L_1(s_k, a_k)$ and the indicator $\kappa_1$ over the first stage of training under Algorithm \ref{alg:training_1}, where, to account for randomness, the experiment is repeated 10 times with 10 different policies trained under different randomized reference trajectories.
It can be seen that the RL cost decreases as the learning continues, with in particular the infeasibility indicator approaching zero.
Figure \ref{fig:training_traj} shows the first and last 500 time steps during the first training stage.
Initially, with an untrained policy and heavy exploration, the gear-shift schedule suggested by $\varpi_\theta$ is often infeasible.
The gear choices are erratic, causing in turn erratic tracking behavior as the MPC controller adjusts to the provided gear-shift schedules.
By the end of training infeasibility of the gear-shift schedule suggested by $\varpi_\theta$ is very rare (only one instance in the last 500 time steps) and the tracking behavior is improved.

\begin{figure}[t]
	\centering
	\input{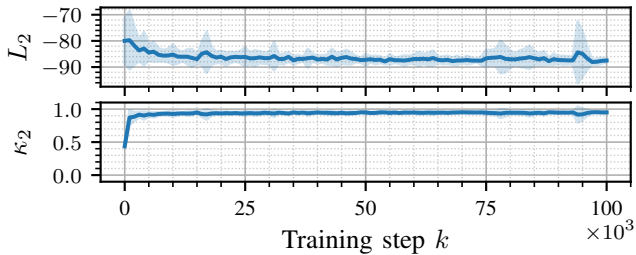}
	\vspace{-0.6cm}
	\caption{Performance metrics over the training stage 2. The plots show mean (solid) and standard deviation (shaded) of the moving average of the metrics for a moving window of 1000 steps.}
	\label{fig:train_stage_2}
\end{figure}

Figure \ref{fig:train_stage_2} shows the costs $L_2(s_k, a_k)$ and the indicator $\kappa_2$, where again, to account for randomness, 10 policies are trained under different random trajectories.
It can be seen that at this second stage the policy rapidly learns the interaction with the heuristic gear controllers.
Indeed, the indicator $\kappa_2$ rapidly approaches an average of one, indicating that the policy learns to improve over the heuristic policies more frequently. 

In the following, we refer to the proposed approach where $\varpi_\theta$ is trained only with the first stage as LC-1, and as LC-2 when trained with both stages.

\subsection{Evaluation}
To evaluate the performance of the controllers we evaluate the performance metric $J(K)$ defined in \eqref{eq:metric}, for $1000$s ($>$ 15 minutes of driving), i.e., $J(1000)$.
The evaluation is performed on 25 random reference trajectories that were not present during training.

The solution to the MINLP \eqref{eq:MINLP} is used as a baseline, from which we define a relative cost increase as 
\begin{equation}
	\Delta J_\mathrm{type}(K) = 100 \cdot \frac{J_\mathrm{type}(K) - J_\text{MINLP}(K)}{J_\text{MINLP}(K)}
\end{equation}
with $J_\text{MINLP}(K)$ the performance metric under MINLP and $J_\mathrm{type}(K)$ the performance metric under the respective other controller, with type$\in$\{MINLP, MIQP, HD, HC, HS, LC-1, LC-2\}.

\definecolor{myblue}{RGB}{95, 158, 160}
\definecolor{myred}{RGB}{255, 71, 3}
\begin{figure*}[ht]
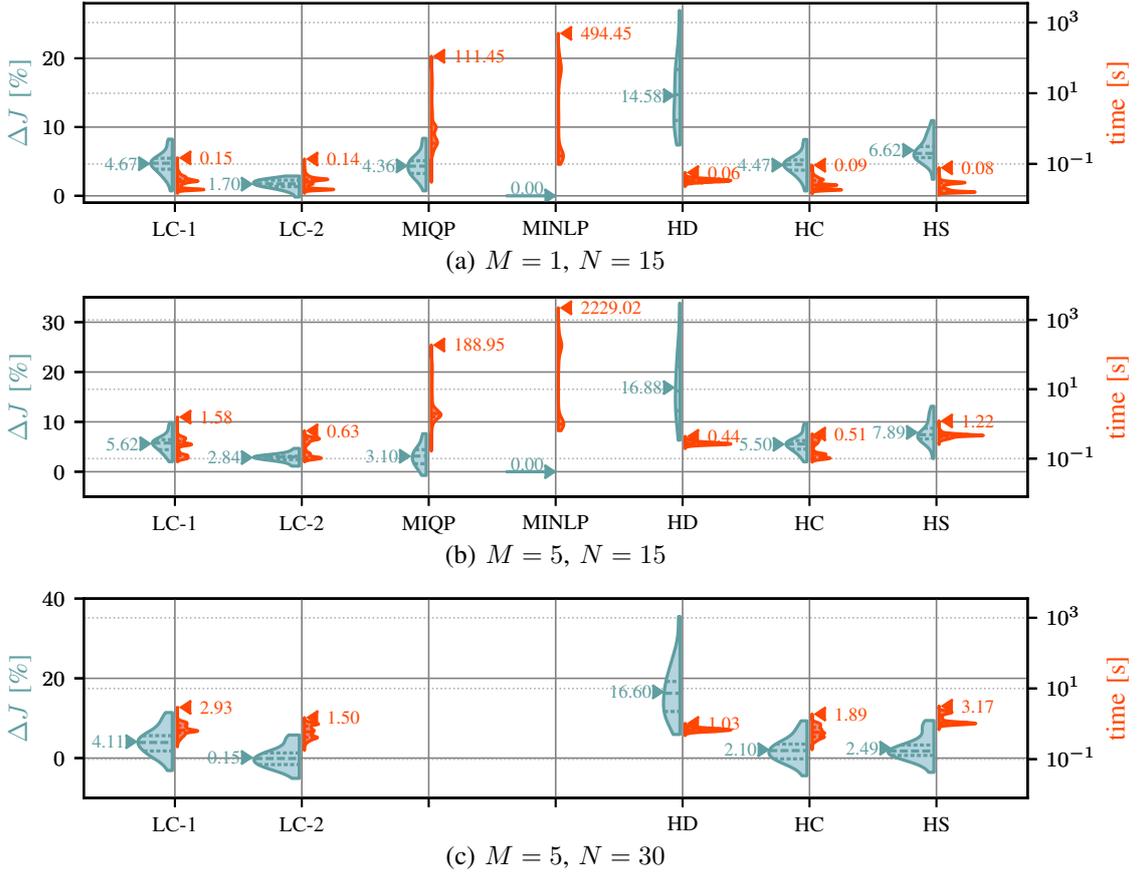

	\centering
	\subfloat{
		\input{media/pgf/eval_single_no_tl.pgf}
	} \\
	\vspace{-0.6cm}
	\subfloat{
		\input{media/pgf/eval_platoon_15.pgf}
	} \\
	\vspace{-0.6cm}
	\subfloat{
		\input{media/pgf/eval_platoon_30.pgf}
	}
	\vspace{-0.3cm}
	\caption{Evaluation of the controllers. The violin plots show the relative performance drop $\Delta J$ (\textcolor{myblue}{blue}, left) and computation time per timestep (\textcolor{myred}{red}, right) of the different controllers evaluated on 25 experiments of 1000 time steps. Note that in (c) the relative performance drop is calculated with respect to the MINLP approach with $M=5, N=15$.}
	\label{fig:eval}
\end{figure*}

\begin{figure}
	\centering
	\subfloat{\input{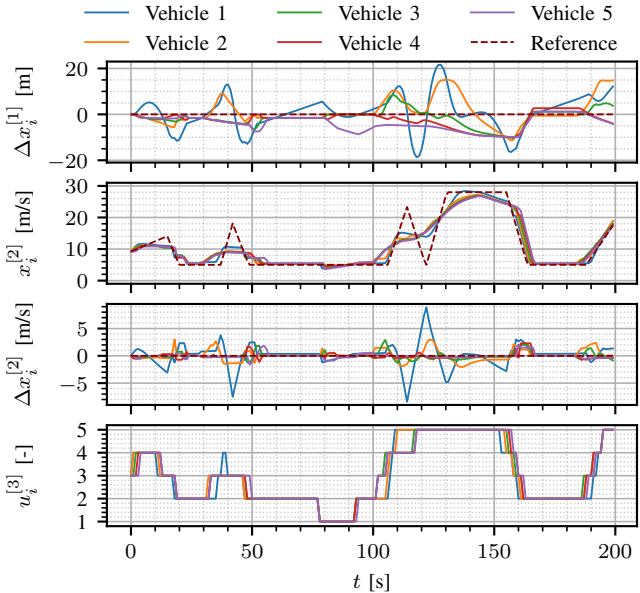}}
	\vspace{-0.3cm}
	\caption{Trajectories and tracking error of the vehicles of a platoon with $M=5$ over 200 time steps.}
	\label{fig:platoon_trajectories}
\end{figure}

Figure \ref{fig:eval}a shows the distributions of the relative cost increase and the solve times for $M=1$, i.e., the single vehicle conditions under which the proposed approach is trained, while Figure \ref{fig:eval}b shows the these results for a platoon with $M=5$.
The advantage of co-optimization is clearly demonstrated by the large cost increase of the HD controller.
Of the co-optimization approaches, MIQP clearly introduces suboptimality through the convexifying approximations.
Interestingly, HC has a comparable performance to MIQP, indicating that the suboptimality of the convexifications is approximately equivalent to that of using a trivial gear sequence with accurate fuel and dynamic models.
Of the newly proposed approaches, LC-1 only slightly improves on the performance of HC controller.
In contrast, LC-2 is significantly higher performing, highlighting the importance of the refinement learning stage 2. 
Indeed, LC-2 has performance very close to the baseline MINLP.
While the baseline is clearly not the truly optimal controller due to the time limit applied (which is highlighted by LC-2 outperforming MINLP in at least one instance), this result demonstrates the proposed approach's superior performance. 
For computation time, it can be seen that solving NLPs online offers a significant computational advantage over mixed-integer approaches.
Notably, LC-2 offers a one or two order of magnitude speed increase over the mixed integer approaches, with negligible suboptimality.
\textcolor{black}{Furthermore, the similar trends in computation and performance between $M=1$ and $M=5$ corroborate how the gear-shift schedule policy $\varpi_\theta$, while trained in a single AV scenario, generalizes to the platoon case.}
Finally, Figure \ref{fig:eval}c presents the same performance indexes for the platoon with an MPC horizon of $N=30$.
Note that both MIQP and MINLP are omitted, as finding feasible solutions to the MPC problems required an unreasonable amount of time\footnote{Further results on the computation time of mixed-integer approaches are reported in the appendix of the extended online version of this article} (more than two days per 1000 time steps).
The ability of $\varpi_\theta$ to generalize over horizons is demonstrated, with the performance of the LC-2 approach improving with the increased horizon, and once again introducing the least suboptimality.
The  cost increase results of Figure \ref{fig:eval} are additionally reported in Table \ref{tab:cost_m_1}, where additional information on the $\Delta J$ distribution is reported.
The values in the table confirm the superior performances of the LC-2 controller to the other methods, and highlights how the proposed approach has a significantly lower standard deviation, indicating a more consistent performance, which is also confirmed by the worst-case performance (`Max' column), which holds a lower value than those of the other controllers. 
Finally, a representative example of the vehicle trajectories with $M=5$, using the proposed approach LC-2 is shown in Figure \ref{fig:platoon_trajectories}.

\begin{table}[t]
	\centering
	\footnotesize
	\caption{$\Delta J$ [\%] statistics over 25 evaluation experiments.}
	\label{tab:cost_m_1}
	\setlength{\tabcolsep}{5pt}
	\begin{tabular}{clrrrrr}
		\toprule
		& \multicolumn{1}{c}{\textbf{Controller}}
		& \multicolumn{1}{c}{\textbf{Mean}}
		& \multicolumn{1}{c}{$\bm{\sigma}$}
		& \multicolumn{1}{c}{\textbf{Median}}
		& \multicolumn{1}{c}{\textbf{Min}}
		& \multicolumn{1}{c}{\textbf{Max}} \\ 
		\midrule
		\multirow{7}{1em}{\rotatebox[origin=c]{90}{$M=1, N=15$}} 
		& LC-1 & 4.67 & 1.66 & 4.75 & 0.72 & 8.24 \\
		& LC-2 & \textbf{1.70} & \textbf{0.68} & \textbf{1.74} & \textbf{-0.17} & \textbf{2.88} \\
		& MIQP & 4.36 & 1.78 & 4.32 & 0.73 & 8.36 \\
		& MINLP & 0.00 & 0.00 & 0.00 & 0.00 & 0.00 \\
		& HD & 14.58 & 4.87 & 14.73 & 7.39 & 26.94 \\
		& HC & 4.47 & 1.58 & 4.58 & 0.72 & 8.20 \\
		& HS & 6.62 & 1.81 & 6.17 & 2.41 & 10.94 \\
		\midrule
		\multirow{7}{1em}{\rotatebox[origin=c]{90}{$M=5, N=15$}} 
		& LC-1 & 5.62 & 1.75 & 5.76 & 2.01 & 9.85 \\
		& LC-2 & \textbf{2.84} & \textbf{0.78} & \textbf{2.94} & 1.12 & \textbf{4.68} \\
		& MIQP & 3.10 & 2.18 & 3.14 & \textbf{-0.74} & 7.62 \\
		& MINLP & 0.00 & 0.00 & 0.00 & 0.00 & 0.00 \\
		& HD & 16.88 & 6.42 & 16.24 & 6.33 & 33.79 \\
		& HC & 5.50 & 1.71 & 5.55 & 1.99 & 9.76 \\
		& HS & 7.89 & 2.33 & 7.42 & 2.66 & 13.13 \\
		\midrule
		\multirow{6}{0.8em}{\rotatebox[origin=c]{90}{$M=5, N=30$}} 
		\rule{0pt}{0.6em} \\[-0.6em]
		& LC-1 & 4.11 & 3.34 & 3.91 & -3.12 & 11.42 \\
		& LC-2 & \textbf{0.15} & \textbf{2.61} & \textbf{-0.10} & \textbf{-5.08} & \textbf{5.79} \\
		& HD & 16.60 & 6.73 & 16.25 & 5.93 & 35.50 \\
		& HC & 2.10 & 3.06 & 1.89 & -4.42 & 9.34 \\
		& HS & 2.49 & 3.02 & 1.77 & -3.58 & 9.42 \\
		\rule{0pt}{0.6em} \\[-0.6em]
		\bottomrule
	\end{tabular}
\end{table}

\section{Conclusions}
\label{sec:conclusions}
We have proposed a novel learning-based MPC controller for fuel-efficient platooning.
By using reinforcement learning to train a policy that selects the gear-shift schedule over the MPC prediction window, vehicle speed and gear position are optimized for fuel-efficient tracking, without the computational burden of solving a mixed-integer program.
An architecture is proposed that allows a policy trained using single-agent RL to generalized to the multi-agent (vehicle) case.
In numerical simulations the proposed approach is shown to achieve comparable performance to highly computational intensive approaches that solve mixed-integer programs.
Furthermore, the online computational burden is improved by more than two orders of magnitude.

Future work will look at guaranteeing feasibility of the gear-shift schedule policy with hard collision constraints, as well as providing suboptimality guarantees on the closed-loop performance when using the learned policy.

\bibliographystyle{IEEEtran}
\bibliography{bib.bib}
\appendices
\section{Approximate solutions for mixed integer problems}
The extremely long solution times often required for mixed-integer approaches can be attributed to the branch-and-bound procedure, wherein many relaxed problems are solved until an optimality condition is satisfied \cite{gurobi}.
It is reasonable to ask how these approaches perform in closed loop if the solver is terminated early, taking the best available solution found within a time limit.
Indeed, state-of-the-art mixed integer solvers often employ several heuristic methods to generate feasible solutions of reasonable quality early in the solving process. 

\begin{figure}[t]
	\centering
	\input{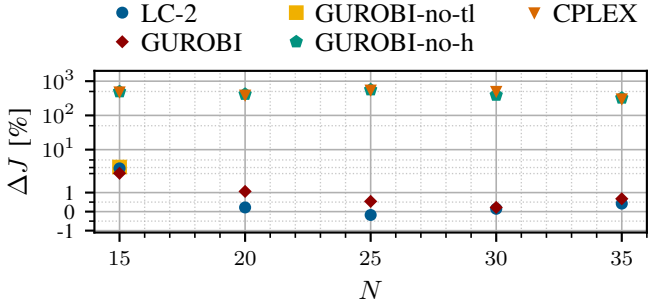}
	\vspace{-0.5cm}
	\caption{Comparison of LC-2 and MIQP controllers over different horizons $N\in\{15, \ldots, 35\}$. The baseline to compute $\Delta J$ is the MINLP controller with $M=5$ and $N=15$.}
	\label{fig:horizon_comparison}
\end{figure}

In the simulations of the current paper, we observed that in particular Gurobi, as a solver of MIQP problems, was able to generate solutions very quickly via heuristic methods that performed relatively well in closed loop.
Figure \ref{fig:horizon_comparison} presents the relative performance drop of the MIQP approach with solve time limited to the maximum required by our proposed approach LC-2 (e.g., 1.5s for $N=30$ as in Figure \ref{fig:eval}), using the commercial solvers Gurobi and CPLEX \cite{cplex2009v12}, compared to that of the proposed approach LC-2.
Furthermore, the results for Gurobi without the use of heuristics (via the `Heuristics' solver parameter) are shown. 
To explore this effect thoroughly, results are shown for horizons $N \in \{15, 20, 25, 30, 35\}$.
Extremely interestingly, with the Gurobi solver, the time-limited MIQP approach outperforms the non-time-limited version of the MIQP approach in Figure \ref{fig:eval}. 
However, for the same controller, using the solver CPLEX results in very poor quality solutions and highly sub-optimal performance.
Furthermore, without heuristics Gurobi also performs very poorly.
We note that CPLEX without time limits performs identically to Gurobi, as is expected due to the global optimum property of MIQP problems.
Clearly, the heuristic procedure in this case uniquely allows the Gurobi solver to find suboptimal solutions that result in strong closed-loop performance.

We attribute this strange behavior of the MIQP approach using the Gurobi solver to a potential overlapping of approximations; the MIQP approach is suboptimal due to model approximations, and in some cases it seems that suboptimal solutions to the optimization problem can then result in improved closed-loop performance.
We note that this behavior is unexpected and cannot be relied upon, as it is clearly solver and situation dependent.
Furthermore, despite this, the proposed approach in general still performs the best.

\section{Details on Comparison Controllers}
Here we provide some additional details on the comparison controllers used in Section \ref{sec:results}.

\subsection{MIQP-based MPC (MIQP)}
This local controller follows the approach from \cite{shao2021vehicle}, where all non-convexities in \eqref{eq:MINLP} are convexified such that the remaining optimization problem is an MIQP.
In particular, the bi-linear term in $J_\text{f}$ is relaxed using a McCormick relaxation \cite{shao2021vehicle}, the quadratic term in the dynamics is replaced by a piecewise-linear approximation, and all bi-linear terms in the dynamics, e.g., $u_1 z(u_3)$, are replaced by mixed-integer inequalities (see \cite{shao2021vehicle} for details).
Convexification introduces errors in the dynamics and the fuel prediction via the piecewise-linear approximation and the McCormick relaxation, respectively; however, in general it lowers the computational burden of this controller compared to the MINLP-based MPC controller.

\subsection{Heuristic Decoupled MPC (HD)}
This local controller follows the principle of decoupling the optimization of the vehicle speed from the gear-shift schedule.
\textcolor{black}{A simplified dynamic model is used within the MPC controller, optimizing the continuous vehicle dynamics, with the gear then selected based on the vehicle velocity.}
To this end, consider the simplified dynamics $x_i(k+1) = \tilde{f}\big(x_i(k), W_i(k)\big)$, where
\begin{equation}
	\label{eq:approx_dynam}
	\tilde{f}(x, W) = \begin{bmatrix}
		x^{[1]} + x^{[2]} \Delta t\\
		x^{[2]} + \frac{\Delta t}{m} (W - C x^{[2]^2} - G)
	\end{bmatrix}.
\end{equation}
The input $W_i$ replaces $T_i z(j_i)z_\text{f}/r - F_i$, the desired braking force and the applied force from the engine torque combined with the gear.
The following NLP is solved:
\begin{subequations}
	\label{eq:heirachical_NLP}
	\begin{align}
		\hat{J}\big(&x_i(k), \hat{\textbf{x}}_i(k), \textbf{p}_i^+(k), \textbf{p}_i^-(k)\big) = \nonumber\\*
		&\hspace{-12pt}\min_{\textbf{x}(k), \textbf{W}_i(k)} \sum_{\tau = 0}^{N} J_\text{t}\big(x_i(\tau|k), \hat{x}_i(\tau+k)\big) \\
		&\quad\quad\quad \quad + \omega\big(\sigma_i^+(\tau|k) + \sigma_i^-(\tau|k)\big) \\
		&\text{s.t.} \quad \eqref{eq:IC_x}, \eqref{eq:acc_lim}, \eqref{eq:collision_ahead}, \eqref{eq:collision_behind}, \eqref{eq:slacks}\\
		&\quad\text{for} \quad \tau = 0,\dots,N-1: \nonumber \\
		&\quad\quad x_i(\tau+1|k) = \tilde{f}\big(x_i(\tau|k), W_i(\tau|k)\big) \\
		&\quad\quad T_\text{min}\frac{z(0) z_\text{f}}{r} - F_\text{max} \leq W_i(\tau|k) \leq W_{i, \text{max}}(k) \\
		&\quad v_\text{min} \leq x_i^{[2]}(\tau|k) \leq v_\text{max} \quad \tau = 0,\dots,N
	\end{align}
\end{subequations}
where the fuel cost cannot be considered as the powertrain dynamics are not modeled.
The bound $W_{i,\text{max}}(k)$ is determined at each time step $k$ by considering the gear that provides the most traction for the current velocity:
\begin{equation}
	W_{i, \text{max}}(k) = T_\text{max} \cdot \max_{j \in \Phi(x_i^{[2]}(k))} \frac{z(j) z_\text{f}}{r}.
\end{equation}
The gear is then selected as $j_i(k\Delta t) = \phi\big(x_i^{[2]}(k)\big)$ and clipped such that \eqref{eq:gear_shift_lim} is respected.
In our simulations we found
\begin{equation}
	\phi(v) = \max_{j\in\Phi(v)} j
\end{equation}
to perform best for this controller, as this corresponds to the available gear that keeps the lowest engine speed, saving on fuel consumption.
Finally, $T_i(k \Delta t)$ and $F_i(k\Delta t)$ are decided as 
\begin{equation}
	\displaystyle
	\begin{aligned}
		&T_i(k\Delta t) = \begin{cases}
			\displaystyle
			T_\text{min} & \text{if } W_i^\ast(0|k) < 0 \\
			\displaystyle
			\frac{W_i^\ast(0|k) r}{z(j_i(k\Delta t)) z_\text{f}} & \text{if } W_i^\ast(0|k) \geq 0
		\end{cases}, \\
		&F_i(k\Delta t) = \begin{cases}
			\displaystyle
			\frac{T_\text{min} z(j_i(k\Delta t)) z_\text{f}}{r} -W_i^\ast(0|k) & \hspace{-7pt} \text{if } W_i^\ast(0|k) < 0 \\
			0 & \hspace{-7pt} \text{if } W_i^\ast(0|k) \geq 0
		\end{cases},
	\end{aligned}
\end{equation}
with the torque rate constraint \eqref{eq:T_lim} applied afterwards with clipping.

\subsection{Heuristic Co-optimization MPC (HC)}
This local controller solves the NLP \eqref{eq:NLP} with the gear-shift schedule chosen by the heuristic $\textbf{j}_i(k) = \rho_{\text{const}, \phi}\big(x_i(k)\big)$.
Similar to a multi-starting approach, the NLP is solved in parallel for several heuristic gear-shift schedules; namely $\rho_{\text{const}, \phi_1}, \rho_{\text{const}, \phi_2},$ and $\rho_{\text{const}, \phi_3}$ corresponding to taking the lowest, highest, and middle of the feasible gear choices, i.e.,
\begin{color}{black}
	\begin{equation}
		\begin{aligned}
			\displaystyle
			\phi_1(v) &= \min_{j\in\Phi(v)} j , \\
			\phi_2(v) &= \max_{j\in\Phi(v)} j,	\\
			\phi_3(v) &= \mathrm{floor}\left(\frac{\phi_2(v) - \phi_1(v)}{2}\right).
		\end{aligned}
		\label{eq:heuristics}
	\end{equation}
\end{color}
The control input is then taken from the minimizer of the optimization problem with the lowest cost.

\subsection{Heuristic Shifted-solution Co-optimization MPC (HS)}
This local controller solves the NLP \eqref{eq:NLP} with the gear-shift schedule obtained by shifting the gear-shift schedule from the previous time step and selecting the last element via $\phi_2$
\begin{equation}
	\textbf{j}_i(k) \!=\! \big[j_i(1|k-1), \dots, j_i(N-1|k-1), \phi_2(x_i^{*, [2]}(N|k-1))\big]^\top.
\end{equation}

\vfill
\end{document}